\documentclass[aps,prd,showpacs,twocolumn,floatfix,amsmath,amssymb,superscriptaddress,preprintnumbers]{revtex4-1}

\usepackage{hyperref}
\usepackage{bm}
\usepackage{graphicx}
\usepackage{color}

\newcommand{\mnras}{Mon. Not. Roy. Astron. Soc.}

\newcommand{\jcap}{J. Cosmol. Astropart. Phys.}
\newcommand{\physrep}{Phys. Rept.}

\newcommand{\pasj}{Publ. Astron. Soc. Jpn.}
\newcommand{\pasp}{Publ. Astron. Soc. Pac.}

\newcommand{\beq}{\begin{equation}}
\newcommand{\eeq}{\end{equation}}
\newcommand{\beqa}{\begin{eqnarray}}
\newcommand{\eeqa}{\end{eqnarray}}
\newcommand{\nwrm}{\mathrm{nw}}
\newcommand{\wrm}{\mathrm{w}}
\newcommand{\Lrm}{\mathrm{L}}
\newcommand{\oneloop}{\text{1-loop}}
\newcommand{\twoloop}{\text{2-loop}}
\newcommand{\sigmad}{\sigma_\mathrm{d}}
\newcommand{\Mpc}{\mathrm{Mpc}}

\newcommand{\hMpc}{h^{-1}\,\mathrm{Mpc}}
\newcommand{\hGpc}{h^{-1}\,\mathrm{Gpc}}
\newcommand{\hMpcinv}{h\,\mathrm{Mpc}^{-1}}

\begin{document}

\preprint{YITP-18-107}

\title{Perturbation theory challenge for cosmological parameters estimation: Matter power spectrum in real space}

\author{Ken Osato}
\email[]{ken.osato@utap.phys.s.u-tokyo.ac.jp}
\affiliation{Department of Physics, School of Science, The University of Tokyo,
Tokyo 113-0033, Japan}
\affiliation{Institut d'Astrophysique de Paris, Sorbonne Universit\'e, CNRS, UMR 7095, 75014 Paris, France}

\author{Takahiro Nishimichi}
\affiliation{Kavli Institute for the Physics and Mathematics of the Universe,
The University of Tokyo Institutes for Advanced Study (UTIAS), The University of Tokyo,
Chiba 277-8583, Japan}

\author{Francis Bernardeau}
\affiliation{Institut d'Astrophysique de Paris, Sorbonne Universit\'e, CNRS, UMR 7095, 75014 Paris, France}
\affiliation{Institut de Physique Th\'eorique, Universit\'e Paris-Saclay, CEA, CNRS, URA 2306,
91191 Gif-sur-Yvette, France}

\author{Atsushi Taruya}
\affiliation{Center for Gravitational Physics, Yukawa Institute for Theoretical Physics,
Kyoto University, Kyoto 606-8502, Japan}
\affiliation{Kavli Institute for the Physics and Mathematics of the Universe,
The University of Tokyo Institutes for Advanced Study (UTIAS), The University of Tokyo,
Chiba 277-8583, Japan}

\date{\today}

\begin{abstract}
We study the accuracy with which cosmological parameters can be determined
from real space power spectrum of matter density contrast
at weakly nonlinear scales using analytical approaches.
From power spectra measured in $N$-body simulations and
using Markov chain Monte-Carlo technique,
the best-fitting cosmological input parameters are determined
with several analytical methods as a theoretical template,
such as the standard perturbation theory, the regularized perturbation theory,
and the effective field theory.
We show that at redshift 1,
all two-loop level calculations can fit the measured power spectrum
down to scales $k \sim 0.2 \, \hMpcinv$ and
cosmological parameters are successfully estimated in an unbiased way.
Introducing the Figure of bias (FoB) and Figure of merit (FoM) parameter,
we determine the validity range of those models
and then evaluate their relative performances.
With one free parameter, namely the damping scale,
the regularized perturbation theory is found to be able to provide
the largest FoM parameter while keeping the FoB in the acceptance range.
\end{abstract}


\maketitle

\section{Introduction}
\label{sec:intro}
The primordial density fluctuations,
which are believed to be generated quantum mechanically
during inflationary stage of the Universe,
has evolved under the influence of cosmic expansion and gravity,
and resulted in rich structures over the observable scale of the Universe.
Thus, the statistical nature of the large-scale structure of the Universe,
as partly traced by galaxy redshift surveys and weak lensing surveys,
contains rich cosmological information.
Given large-scale structure data set,
statistical inference of cosmological parameters
as well as the test of cosmological models are now a routine task,
and increasing the statistical precision,
an efficient and unbiased way to extract the cosmological information is rather critical.
In particular, the baryon acoustic oscillations (BAO)
imprinted on the spatial clustering pattern of galaxies
is known as primeval acoustic signature of the baryon-photon fluid,
and is used as a standard ruler to constrain the late-time cosmic acceleration
through the precise measurement of power spectrum or correlation function \cite{Aubourg15}.
While the BAO is thought to be a robust and idealistic cosmological probe,
several systematics in reality come to play.
Accurately modeling or removing those systematics is currently the central issue
in precision cosmology with large-scale structure observations.

Over the last decade, there have been survey projects which aimed at
detecting BAO via the measurements of galaxy power spectrum, e.g.,
Baryon Oscillation Spectroscopic Survey (BOSS) \cite{Ross17,Beutler17},
6dF Galaxy Survey \cite{Beutler11}, and WiggleZ \cite{Blake11}.
In addition, projects which span larger areas and can detect more objects
have been proposed, for example, Dark Energy Spectroscopic Instrument (DESI) \cite{DESI18},
Subaru Prime Focus Spectrograph (PFS) \cite{PFS14},
Large Synoptic Survey Telescope (LSST) \cite{LSST09}, and
\textit{Euclid} \cite{Laureijs11,Amendola13}.
Since these surveys measure power spectrum at sub-percent level,
we need more accurate and precise modeling of power spectrum
over a wider range of scales to constrain cosmological parameters and
as well as to test cosmological models at high precision.

A standard approach to get accurate power spectrum over the wide range of scales
is numerical simulations.
Among other numerical methods, $N$-body simulation,
in which the smooth matter distribution is described approximately
by the collections of discrete particles,
is a suitable tool to trace the nonlinear gravitational evolution of the matter fluctuation
because nonlinear evolution is efficiently taken into account
down to the scales limited by the resolution.
However, running $N$-body simulations to explore large parameter spaces
is not practical because of the large computational cost.
In many cases, analytical prescriptions are employed to efficiently compute
power spectrum with numerous sets of cosmological parameters in a forward modeling manner
and infer the cosmological parameters from measurements of power spectrum.
For this purpose, several approaches have been proposed
to accurately predict power spectrum.
The perturbation theory (PT) has played a central role to compute power spectrum
analytically (see Ref.~\cite{Bernardeau02} for a comprehensive review).
Under the single-stream approximation, the system to solve
is reduced to the cosmic fluid which follows the continuity, Euler, and
Poisson equations in the expanding Universe, and one can expand these equations
with respect to the linear density contrast.
This naive perturbative approach is called as standard perturbation theory (SPT).
However, it is known that SPT has poor convergence properties
when higher order correction is included,
and deviates from the results of $N$-body simulations
from a relatively large scale \cite{Blas14}.

In order to improve the convergence and
compute power spectrum more reliably
on small scales, approaches alternative to SPT have been presented
based on resummation schemes in Lagrangian \cite{Matsubara08} or
in Eulearian \cite{Crocce06,Bernardeau08,Bernardeau12} space.
Analytical approaches have been further extended in the context of
effective field theories (EFT) \cite{Baumann12,Carrasco12,Carrasco14},
which incorporate small scale physics, beyond the single stream regime, by introducing
effective interaction terms in the equation of motions.
The scales of validity of those models obviously vary from models to models.
Furthermore, important effects, e.g., galaxy bias or redshift space distortion,
can be taken into account in some models \cite{Taruya10,Matsubara14}.
In Ref.~\cite{FonsecadelaBella18}, they investigate how
choice of analytical approaches and models of bias and redshift space distortion
affects the goodness of fit in the case of power spectrum.
The goal we pursue here is to give first insights in the ability of such models to not only
reproduce $N$-body results at one representative cosmological parameter set, but also
to effectively infer cosmological parameters
from matter power spectrum measurements.
For this purpose, we investigate how accurately these methods
can recover the cosmological parameters
from the full shape measurement of matter power spectrum.
More specifically, we first generate initial conditions
with a set of cosmological parameters, and then run $N$-body simulations
to create a matter density field at the late time Universe.
Following analysis similar to that used in observations,
we measure matter power spectrum from the simulations and
fit it with analytical methods based on Markov chain Monte-Carlo (MCMC) technique.
Finally, we can obtain the constraints on cosmological parameters and compare
them with the parameters used to generate the initial condition.
With such a test, we can then infer the scales down to
which the analytical methods can be used
without biasing the retrieved cosmological parameters
--- within the statistical error-bars --- and then
identify the best performing theoretical modeling.
In addition to commonly used methods like SPT,
we employ the following extended theories.
One is the regularized perturbation theory (\texttt{RegPT}) \cite{Taruya12}.
In this framework, the expansion based on SPT
is reorganized based on multipoint propagator expansion.
Other approaches we test in this study are based on EFT constructions
where one or several free parameters are introduced
to describe the impact of the small scale physics,
such as the effective pressure of cosmic fluid,
to the growth of the large scale modes.
A specific goal of our study precisely lies in
investigating the usefulness of such models with free parameters.
The presence of those parameters extends naturally the validity range of models
but require either their calibration in simulations or
their joint fitting, together with the cosmological parameters. In this study
we test these models by fitting the free parameters they contain simultaneously
with the set of cosmological parameters we choose.
Furthermore, for reference we also use the response function approach \cite{Nishimichi16},
which is a simulation-aided approach for computing nonlinear power spectrum.

This paper is organized as follows.
In Sec.~\ref{sec:theory}, we review basics of analytical approaches:
SPT, \texttt{RegPT}, IR-resummed EFT, and the response function approach.
We give details of $N$-body simulations and parameter estimation
in Sec.~\ref{sec:methods}.
Then, we present analysis of parameter estimation with the power spectrum
measured from simulations in Sec.~\ref{sec:results}.
We conclude in Sec.~\ref{sec:conclusions}.

Throughout the paper, we assume a flat $\Lambda$ cold dark matter Universe model,
and fiducial cosmological parameters are as follows:
scaled Hubble parameter $h = H_0/(100 \, \mathrm{km} \, \mathrm{s}^{-1} \, \Mpc^{-1}) = 0.6727$,
physical cold dark matter density $\Omega_\mathrm{c} h^2 = 0.1198$,
baryon density $\Omega_\mathrm{b} h^2 = 0.02225$,
the amplitude $A_\mathrm{s} = 2.2065 \times 10^{-9}$ and
the tilt $n_\mathrm{s} = 0.9645$ of scalar perturbation
at the pivot scale $k_\mathrm{piv} = 0.05 \, \Mpc^{-1}$.
Then, the total matter density $\Omega_\mathrm{m}$ is the sum of dark matter,
baryon and massive neutrino components
and we assume neutrinos compose of two massless neutrinos
and one massive neutrino with the mass $M_\nu = 0.06 \, \mathrm{eV}$,
which corresponds to the physical energy density $\Omega_\nu h^2 = 0.00064$.
Note that the effect of massive neutrinos are taken into account only in the computation of
linear matter power spectrum at $z = 0$.
Both the simulations and the analytical calculations
are based on the linear power spectrum scaled to
the initial redshift or the redshift at which the nonlinear spectra are
computed assuming a scale-independent linear growth factor ignoring the masses of neutrinos.

\section{Theory}
\label{sec:theory}
In this Section, we briefly review several PT approaches to
analytically compute matter power spectrum on
weakly nonlinear scales.

\subsection{Standard Perturbation Theory}
In this prescription, we begin with fluid equations in the single-stream
approximation (continuity, Euler, and Poisson equations)
and then the density and velocity fields are
expanded with respect to the linear density contrast.
It is useful to expand the fields in Fourier space
because it clarifies how the mode couples with each other.
The resultant expansion of the density field $\delta$ is expressed in powers of
linear density field $\delta_0$ at the present Universe,
\beqa
\delta (\bm{k}) &=& \sum_{n=1} D_+^n \delta^{(n)} (\bm{k}),
\label{eq:SPT1} \\
\delta^{(n)} (\bm{k}) &=& \int \frac{d^3 q_1 \cdots d^3 q_n}{(2 \pi)^{3(n-1)}}
\delta_\mathrm{D} (\bm{k} - \bm{q}_1 - \cdots - \bm{q}_n) \nonumber \\
& & \times F_\mathrm{sym}^{(n)}(\bm{q}_1 , \ldots , \bm{q}_n) \delta_0 (\bm{q}_1)
\cdots \delta_0 (\bm{q}_n) ,
\label{eq:SPT2}
\eeqa
where $\delta_\mathrm{D}$ is the Dirac delta function,
$D_+$ is the linear growth factor, and
$F_\mathrm{sym}^{(n)}$ is the $n$-th order symmetrized kernel,
which characterizes mode coupling via the non-linear evolution.
The kernels can be analytically constructed \cite{Goroff86,Crocce06}.

A key statistical property for a statistically homogeneous stochastic
density field is its power spectrum $P(k)$ defined as,
\beq
\langle \delta (\bm{k}) \delta (\bm{k}') \rangle
\equiv (2 \pi)^3 \delta_\mathrm{D} (\bm{k}+\bm{k}') P(k) .
\eeq
The power spectrum depends only on the magnitude of $\bm{k}$
due to the statistical isotropy.
Assuming that the linear density field follows the Gaussian statistics, and using
the perturbative expansion in Eqs.~\eqref{eq:SPT1} and \eqref{eq:SPT2},
we can express the power spectrum perturbatively,
\beq
P(k) = D_+^2 P_0 (k) + \Delta P^\mathrm{SPT}_\oneloop (k) +
\Delta P^\mathrm{SPT}_\twoloop (k) + \cdots ,
\label{eq:SPT_power}
\eeq
where $P_0 (k)$ is linear power spectrum defined as
\beq
\langle \delta_0 (\bm{k}) \delta_0 (\bm{k}') \rangle
\equiv (2 \pi)^3 \delta_\mathrm{D} (\bm{k}+\bm{k}') P_0 (k) .
\eeq
The first and second terms in Eq.~\eqref{eq:SPT_power}
are called as 1-loop and 2-loop correction terms
which involve square and cubic powers of linear power spectrum, respectively.
The explicit expressions for correction terms can be found
in Appendix~\ref{sec:ex_SPT}.

\subsection{Regularized Perturbation Theory}
As an extended PT treatment that improves the convergence of PT expansion
by reorganizing the infinite series of SPT expansion,
we consider a model based on multipoint propagator expansion,
\texttt{RegPT} \cite{Taruya12}.
Here, we review the basic formalism of density power spectrum
according to this framework.

First, we construct $(n+1)$-point propagator $\Gamma^{(n)}$ as
an ensemble average of functional derivatives,
\beqa
&& \frac{1}{n!} \left\langle \frac{\delta^n \delta (\bm{k}, \eta)}
{\delta \delta_0 (\bm{k}_1)\cdots \delta \delta_0 (\bm{k}_n)}
\right\rangle \equiv \nonumber \\
&& \delta_\mathrm{D} (\bm{k} - \bm{k}_{1 \ldots n})
\frac{1}{(2 \pi)^{3(n-1)}} \Gamma^{(n)} (\bm{k}_1 , \cdots , \bm{k}_n),
\eeqa
where $\bm{k}_{1 \ldots n} = \bm{k}_1 + \cdots + \bm{k}_n$,

\begin{widetext}
\beqa
\Gamma^{(n)} (\bm{k}_1, \ldots, \bm{k}_n) &=&
D_+^n F^{(n)}_{\mathrm{sym}} (\bm{k}_1, \ldots, \bm{k}_n) +
\sum_{p=1}^{\infty} \Gamma^{(n)}_{p\text{-loop}}
(\bm{k}_1, \ldots, \bm{k}_n) , \\
\Gamma^{(n)}_{p\text{-loop}} (\bm{k}_1, \ldots, \bm{k}_n) &=&
D_+^{n+2p} c_p^{(n)} \int \frac{d^3 q_1 \cdots d^3 q_p}{(2\pi)^{3p}}
F^{(n+2p)}_{\mathrm{sym}} (\bm{q}_1, -\bm{q}_1, \ldots, \bm{q}_p, -\bm{q}_p,
\bm{k}_1, \ldots, \bm{k}_n) P_0 (q_1) \cdots P_0 (q_p) ,
\label{eq:gamma_propagator} \\
P^{\texttt{RegPT}} (k) &=& \sum_{n=1}^{\infty} n!
\int \frac{d^3 q_1 \cdots d^3 q_n}{(2\pi)^{3(n-1)}}
\delta_\mathrm{D} (\bm{k} - \bm{q}_{1 \ldots n})
[ \Gamma^{(n)} (\bm{q}_1, \ldots, \bm{q}_n) ]^2 P_0 (q_1) \cdots P_0 (q_p) ,
\eeqa
\end{widetext}
where $c_p^{(n)} = {}_{(n+2p)}C_{n} (2p-1)!!$ and
${}_{(n+2p)}C_{n}$ is the binomial coefficient.

The propagator $\Gamma^{(n)}$ has an asymptotic form in high-$k$ limit.
It is shown that \cite{Bernardeau12}
\beqa
\lim_{k \to \infty} \Gamma^{(n)} (\bm{k}_1, \ldots, \bm{k}_n) &=&
\exp \left( - \frac{k^2 D_+^2 \sigmad^2}{2} \right) \nonumber \\
& & \times \Gamma^{(n)}_\mathrm{tree} (\bm{k}_1, \ldots, \bm{k}_n) ,
\label{eq:gamma_highk}
\eeqa
where $k = k_{1 \ldots n}$.
Here, the tree term $\Gamma^{(n)}_\mathrm{tree}$ is identical to
the SPT kernel $D_+^n F^{(n)}_\mathrm{sym}$, and $\sigmad^2$ is
the root-mean-square of one-dimensional displacement field,
\beqa
\sigmad^2 &\equiv& \frac{1}{3} \int \frac{d^3 q}{(2 \pi)^3}
\frac{P_0 (q)}{q^2} \nonumber \\
&=& \int \frac{d q}{6 \pi^2} P_0 (q).
\eeqa
This quantity controls the damping behavior in high-$k$ regime and
is sensitive to integration range.
Ref.~\cite{Taruya12} proposed the running UV cutoff to reproduce
the spectra measured from $N$-body simulations,
\beq
\sigmad^2 (k) = \int_0^{k_\Lambda (k)} \frac{d q}{6 \pi^2} P_0 (q) ,
\eeq
where the UV cutoff scale is $k_\Lambda (k) = k/2$.

Then one can construct the \textit{regularized} propagators
which approaches toward the expected asymptotes at both ends,
Eq.~\eqref{eq:gamma_highk} at the high-$k$ limit and
SPT at the low-$k$ limit.
The expressions for 2-loop and 1-loop levels are found
in Appendix~\ref{sec:ex_RegPT}.
The damping factor is crucial to determine the shape on
high-$k$ regime. In addition to the running UV cutoff case
employed in the original \texttt{RegPT} code,
we investigate a simple generalization of this model by treating $\sigmad$ as a free parameter.
In what follows, we call this version as \texttt{RegPT+}.

\subsection{IR-resummed Effective Field Theory}
The EFT of the large-scale structure provides a way to
incorporate the effects of small-scale physics, beyond shell-crossing, by introducing counter terms.
It has been known that once parameters are calibrated with $N$-body simulations,
one can reproduce the measured power spectrum by sub-percent level
up to high-$k$ ($\lesssim 0.30 \, \hMpcinv$).
However, those parameters have explicit cosmological dependence,
and they must be in general treated as free parameters
in practical analysis of cosmological parameter estimation.
In the present study, we examine a simplified treatment of IR-resummed EFT as presented in
Refs.~\cite{Baldauf15,Vlah16},
where the damping of the BAO wiggle feature in the power spectrum
due to the large-scale bulk motion is modeled by the so-called IR-resummation.

First, we split the power spectrum into the smooth and wiggle part.
For the linear power spectrum, the smoothed part
is evaluated using a featureless transfer function as
\beq
P_\Lrm^\nwrm (k) = P_\mathrm{EH} (k) \mathcal{F} [P_\Lrm (k) / P_\mathrm{EH} (k)] ,
\label{eq:smoothing1}
\eeq
where $P_\Lrm (k)$ is linear power spectrum at a given redshift, i.e.,
$P_\Lrm (k) = D_+^2 P_0 (k)$, and
$P_\mathrm{EH} (k)$ is the power spectrum
from the no-wiggle formula given by Ref.~\cite{Eisenstein98}.
The functional $\mathcal{F} [f(k)]$ represents a smoothing operation defined as,
\beqa
\mathcal{F} [f(k)] &=& \frac{1}{\sqrt{2\pi} \log_{10} \lambda} \int d(\log_{10} q) f(q)
\nonumber \\
&& \times \exp \left[ -\frac{(\log_{10} k - \log_{10} q)^2}{2(\log_{10} \lambda)^2} \right] ,
\label{eq:smoothing2}
\eeqa
where $\lambda$ determines the smoothing scale and
we adopt $\lambda = 10^{0.25} \, \hMpcinv$ \cite{Vlah16}.
That is, we adjust the slight difference in the broadband between the formula by \cite{Eisenstein98}
and the linear power spectrum computed by \texttt{CAMB}, and obtain a smooth baseline model that traces
the overall shape of the linear power spectrum better than $P_\mathrm{EH}(k)$.
The wiggle part $P_\Lrm^\wrm$ is obtained
by subtracting the smooth part $P_\Lrm^\nwrm$ from the total spectrum $P_\Lrm$.
The higher order correction terms, i.e., $\Delta P_\oneloop^\nwrm$ and $\Delta P_\twoloop^\nwrm$,
are obtained by plugging the no-wiggle linear spectrum $P_\Lrm^\nwrm$
into SPT formulas instead of linear power spectrum $P_\Lrm$.
Similarly, the wiggle terms $\Delta P_\oneloop^\wrm$ and $\Delta P_\twoloop^\wrm$
are obtained as the residuals.
Finally, we can compute the matter power spectrum based on IR-resummed EFT
at 2-loop level as,
\begin{widetext}
\beqa
P^\text{IR EFT}_\twoloop (k) &=& P^\nwrm (k) + P^\wrm (k) ; \\
P^\nwrm (k) &=& (1+\alpha_1 k^2) P_\Lrm^\nwrm (k) + (1+\alpha_2 k^2) \Delta P_\oneloop^\nwrm (k)
+ \Delta P^\nwrm_\twoloop (k) , \\
P^\wrm (k) &=& e^{-k^2 \Sigma^2} \left[ (1+\alpha_1 k^2 + C_1) P_\Lrm^\wrm (k)
+ (1+\alpha_2 k^2 + C_2) \Delta P_\oneloop^\wrm (k) + \Delta P_\twoloop^\wrm (k) \right],
\eeqa
\end{widetext}
where
\beqa
C_1 &=& k^2 \Sigma^2 (1+\alpha_1 k^2) + \frac{1}{2} k^4 \Sigma^4 , \\
C_2 &=& k^2 \Sigma^2 (1+\alpha_2 k^2) .
\eeqa
Here we introduce three free parameters, $\alpha_1$, $\alpha_2$, and $\Sigma$.
The parameters $\alpha_1$ and $\alpha_2$ roughly correspond to the effective sound speed
and $\Sigma$ controls the damping behavior at small scales.
The explicit formulas are also found in Appendix~\ref{sec:ex_IRresum}.

\section{Methods}
\label{sec:methods}
In order to test the analytical treatments presented in the previous section,
we conduct a mock cosmological analysis.
First, we generate initial conditions with a given set of cosmological parameters
and then run $N$-body simulations to obtain matter distribution at the late-time Universe.
With the simulated power spectrum and theoretical approaches,
we infer cosmological parameters and compare them with the true values,
i.e., those used to generate initial conditions.

\subsection{$N$-body simulations}
In order to carry out the cosmological parameter estimation,
we need a measured data of matter power spectrum,
for which we use $N$-body simulations.
We run $N$-body simulations to obtain the matter distribution
at the redshift $z = 1$.
We employ $2048^3$ particles and the length on a side is $2 \, \hGpc$.
The initial conditions (ICs) are generated at the redshift $z = 28.683$.
The standard way to create ICs is generating them as Gaussian random field
according to linear power spectrum.
However, this IC is subject to the large sample variance at low-$k$ regime,
which might affect the parameter estimation.
In order to circumvent this effect and improve the convergence,
we employ suppressed variance initial conditions \cite{Angulo16},
in which the norm of Fourier modes, $|\delta_0 (\bm{k})|$, is fixed to its expectation value
from linear power spectrum and then two simulations with inverted phases are paired.
Since the fluctuations in the measured power spectra are partly cancelled by taking the mean of the pair,
this IC can greatly reduce the variance.
Then, we simulate the gravitational evolution of the matter distribution
with the Tree-PM code \texttt{Gadget-2} \cite{Springel05}.
Finally, we measure power spectrum from
the particle distribution with fast Fourier transform.
Our simulation template is based on the average over $5$ pairs of simulations,
and the typical statistical error is sub-percent level.
The input cosmological parameters used to generate the IC are
$h = 0.6727$, $\Omega_\mathrm{c} h^2 = 0.1198$,
$\Omega_\mathrm{b} h^2 = 0.02225$, $\Omega_\nu h^2 = 0.00064$,
$n_\mathrm{s} = 0.9645$, and $A_\mathrm{s} = 2.2065 \times 10^{-9}$.
The derived total matter density parameter is
$\Omega_\mathrm{m} = \Omega_\mathrm{c} + \Omega_\mathrm{b} + \Omega_\nu = 0.3153$.

\subsection{Inference of input cosmological parameters from power spectrum}
\label{sec:estimation}
Here, we estimate cosmological parameters with the methods
described in Sec.~\ref{sec:theory}
and the power spectrum measured from simulations.
To summarize, we consider 4 analytical approaches:
\texttt{RegPT}, SPT, \texttt{RegPT+}, and IR-resummed EFT.

The likelihood distribution $\mathcal{L} (\bm{\theta}|\hat{P})$
is given as the form of multivariate Gaussian distribution,
\begin{widetext}
\beq
\mathcal{L} (\bm{\theta}|\hat{P}) =
\frac{1}{\sqrt{(2\pi)^n \det C }}
\exp \left[ -\frac{1}{2} \sum_{i, j}^n
(\hat{P}(k_i) - P(k_i;\bm{\theta}))
(C^{-1})_{ij} (\hat{P}(k_j) - P(k_j;\bm{\theta})) \right],
\label{eq:posterior}
\eeq
\end{widetext}
where $C$ is the covariance matrix of power spectrum,
$\hat{P}(k)$ is the measured power spectrum from simulations,
and $P (k;\bm{\theta})$ is the prediction based on analytical schemes
with parameters $\bm{\theta}$, which include cosmological parameters
and also nuisance parameters in cases of \texttt{RegPT+} and IR-resummed EFT.
We consider three dimensional cosmological parameter space of
$(h, \Omega_\mathrm{m}, A_\mathrm{s})$
to get converged results within reasonable time.
The parameter $A_\mathrm{s}$ directly determines
the amplitude of matter power spectrum and
varying parameters $h$ or $\Omega_\mathrm{m}$ changes the distance scales
at large scale.
Furthermore, large $\Omega_\mathrm{m}$ enhances the nonlinear growth of
matter fluctuations at small scales.
The measurement of matter power spectrum is known to give
tight constraints on these parameters and
that is why we focus on these parameters in this study.
When varying $\Omega_\mathrm{m}$, we fix the ratio between matter and baryon density
$\Omega_\mathrm{b} / \Omega_\mathrm{m} = 0.1559$.
The information about cosmological parameters is summarized in Table~\ref{tab:values}.
We adopt flat prior for all parameters and varied parameters have reasonable ranges
(see Table~\ref{tab:values}). The prior is zero outside the ranges.
The setting of binning of wavenumbers, i.e.,
the minimum $k_\mathrm{min}$, maximum $k_\mathrm{max}$, and the interval $\Delta k$,
is shown in Table~\ref{tab:kbin}.
Note that all bins are linearly spaced.

For the covariance matrix,
we consider only the Gaussian part
along with the shot noise contribution,
which is given by
\beq
C_{ij} = \frac{2}{N_{k_i}} \left[ P(k_i) + \frac{1}{n_\mathrm{gal,eff}} \right]^2
\delta_{ij},
\eeq
where we define the effective number density
$n_\mathrm{gal,eff} \equiv b_g^2 n_\mathrm{gal}$
with galaxy bias $b_g$ and the number density of galaxies $n_\mathrm{gal}$,
$N_{k_i}$ is the number of the mode, and $\delta_{ij}$ is the Kronecker delta.
Strictly speaking, due to mode coupling,
off-diagonal terms appear in the covariance matrix.
However, the impacts by these terms can be ignored on our interested scales
\cite{Takahashi2009}.
Note that the galaxy bias is introduced only to adjust
the relative contribution of the shot noise to the survey setting that we consider.
The matter power spectrum, not galaxy power spectrum,
is considered throughout the analysis.
The shot noise term regulates the available scales,
where information can be extracted.
Otherwise, the constraints on parameters are determined
only from power spectrum on small scales.
We count the number of modes $N_k$ in the simulations
where the periodic boundary condition is adopted.
In our case, we adopt the survey volume $V = 8.0 \, (\hGpc)^3$,
the galaxy bias $b_g = 1.41$, and the galaxy number density
$n_\mathrm{gal} = 8.4 \times 10^{-4} \, (\hMpc)^{-3}$,
which gives the effective number density $n_\mathrm{gal,eff} = 1.67 \times 10^{-3} \, (\hMpc)^{-3}$.
These parameters are chosen to match the \textit{Euclid} survey
in the specific redshift bin $0.9 < z < 1.1$ \cite{Amendola13}.

We use an Affine invariant Markov chain sampler \texttt{emcee} \cite{Foreman-Mackey13}
to obtain the posterior distribution.
For the burn-in process, we compute the auto-correlation time $t_c$
of parameters for each chain
and discard the first $2 t_c$ steps.
For convergence, the sampler is run until the total steps is
$50$ times larger than auto-correlation time for all parameters
\footnote{Note that the standard Gelman-Rubin statistic is not suitable
for \texttt{emcee} because this algorithm uses information of
different chains in the ensemble to propose next positions.
The obtained chains are correlated and the resultant Gelman-Rubin statistic
will be underestimated.}.

\begin{table*}
\caption{Cosmological parameters}
\begin{ruledtabular}
\begin{tabular}{cclc}
  \multicolumn{4}{c}{Varied parameters} \\
  Symbol & Value & Explanation & Range \\
  \hline
  $h$ & $0.6727$ & Hubble parameter in the unit of $100 \, \mathrm{km} \, \mathrm{s}^{-1} \, \Mpc^{-1}$ & $0.3 < h < 1.3$\\
  $\Omega_\mathrm{m}$ & $0.3153$ & The matter density at the present Universe &
  $0.01 < \Omega_\mathrm{m} h^2 < 0.99$ \\
  $A_\mathrm{s}$ & $2.2065 \times 10^{-9}$ & The amplitude of scalar perturbation
  at the scale $k_\mathrm{piv} = 0.05 \, \Mpc^{-1}$ &
  $A_\mathrm{s} > 0$ \\
  \hline
  \multicolumn{4}{c}{Fixed parameters} \\
  Symbol & Value & Explanation & Range \\
  \hline
  $\Omega_\mathrm{b}/\Omega_\mathrm{m}$ & $0.1559$ & The baryon density fraction
  & $0.005 < \Omega_\mathrm{b} h^2 < 0.1$ \\
  $n_\mathrm{s}$ & $0.9645$ & The tilt of scalar perturbation & \\
  $\Omega_\nu$ & $0.00064$ & The energy density of massive neutrino & \\
\end{tabular}
\end{ruledtabular}
\label{tab:values}
\end{table*}

\begin{table*}
\caption{Binning of wavenumbers}
\begin{ruledtabular}
\begin{tabular}{cccc}
  Model & $k_\mathrm{min} \, [\hMpcinv]$ & $k_\mathrm{max} \, [\hMpcinv]$ & $\Delta k \, [\hMpcinv]$ \\
  \hline
  \texttt{RegPT} and SPT & $0.004$ & $[0.15,0.18,0.21,0.24,0.27,0.30]$ & $0.004$ \\
  \texttt{RegPT+}, IR-resummed EFT, and \texttt{RESPRESSO} & $0.004$ & $[0.15,0.18,0.21,0.24,0.27,0.30,0.33,0.36]$ & $0.004$ \\
\end{tabular}
\end{ruledtabular}
\label{tab:kbin}
\end{table*}

\subsection{Response function approach}
In order to validate our whole procedure we consider a hybrid approach
recently proposed in Refs.~\cite{Nishimichi16,Nishimichi17},
which, by construction, gives unbiased estimates of the parameters with
computable error bars.
In this approach, nonlinear matter power spectrum is expanded by linear power spectrum,
instead of the linear density contrast,
around a fiducial cosmology at which an accurate simulation data is available.
We then make use of the response function which
describes the way the nonlinear power spectrum responds to
the variation of the linear power spectrum:
\beq
K(k, q) = q \frac{\delta P(k)}{\delta P_\Lrm (q)} .
\label{eq:response}
\eeq
The function $K(k, q)$ was studied with numerical simulations
and perturbation theory in detail in Ref.~\cite{Nishimichi16} and it was found that
while the function is well described by SPT at linear to weakly nonlinear scale (in wavenumber $q$),
it exhibits a strong damping behavior at large $q$, and this is even true
when the other wavenumber $k$ stays in a very large scale.
Then, the follow-up paper \cite{Nishimichi17} presented an analytical model
based on a regularized PT and SPT with damping tail motivated by results of numerical simulations,
which performs well over a wide dynamical range in $q$ for a given $k$ in the mildly nonlinear regime.

Once a reasonable model for the response function $K(k, q)$ is constructed,
one can compute the nonlinear matter power spectrum via
\beq
P (k;\bm{\theta}) = P (k;\bm{\theta}_0) + \int d \ln q \, K(k, q)
\left[ P_\Lrm (q;\bm{\theta}) - P_\Lrm(q;\bm{\theta}_0) \right],
\eeq
where we denote by $\bm{\theta}$ the cosmological parameters in the target cosmology and
by $\bm{\theta}_0$ those in the fiducial cosmological model
in which the simulation data for the nonlinear power spectrum is available.
The \texttt{RESPRESSO} code developed in Ref.~\cite{Nishimichi17} follows this equation
to compute power spectrum in the target cosmology.
Eq.~\eqref{eq:response} is valid as long as the difference in the two linear power spectra is small,
but the code takes into account possible higher order corrections
by considering multiple-step reconstruction with an appropriate cosmology-dependence
in the $K(k, q)$ function, when the target cosmology is quite far from the fiducial
one~\footnote{We do not repeat this here
but the performance of the model against rather extreme cosmological models,
such as $\Omega_\mathrm{m} = 0.15$ or $0.45$ can be found in Ref.~\cite{Nishimichi17}.}.

In this paper, the power spectrum template for the fiducial cosmology,
i.e., the first term in the right-hand-side of Eq.~\eqref{eq:response},
is the same as the simulation data used in this study.
This model should therefore provide, by construction,
an unbiased estimate of the cosmological parameters
when fitted to the simulation data,
no matter up to what wavenumber is considered in the fitting.

The response function approach provides also a natural way
to estimate the Fisher matrix forecast as will be discussed in Sec.~\ref{sec:merit}.
We thus employ this model to discuss the consistency between the MCMC analysis
and the Fisher matrix forecast.
Also, the model provides the best-case scenario for the figure of merit assessment,
where no nuisance parameter is introduced and
the best-fit parameters are unbiased for the reason discussed above.
As a consequence, this model can be used to validate our numerical procedure
and be used as referential perfomance for the analytical approaches
we use in the following.

\section{Results}
\label{sec:results}
In this Section, we show results of the parameter inference
based on the various methods presented in Sec.~\ref{sec:theory}.
All of the results presented in the subsequent sections are based on 2-loop level calculations.
For comparison, the results with 1-loop level calculations are presented
in Appendix~\ref{sec:1loop_results}.

\subsection{Fiducial and best-fit power spectra}
\begin{figure}[htbp]
  \centering
  \includegraphics[width=0.45\textwidth]{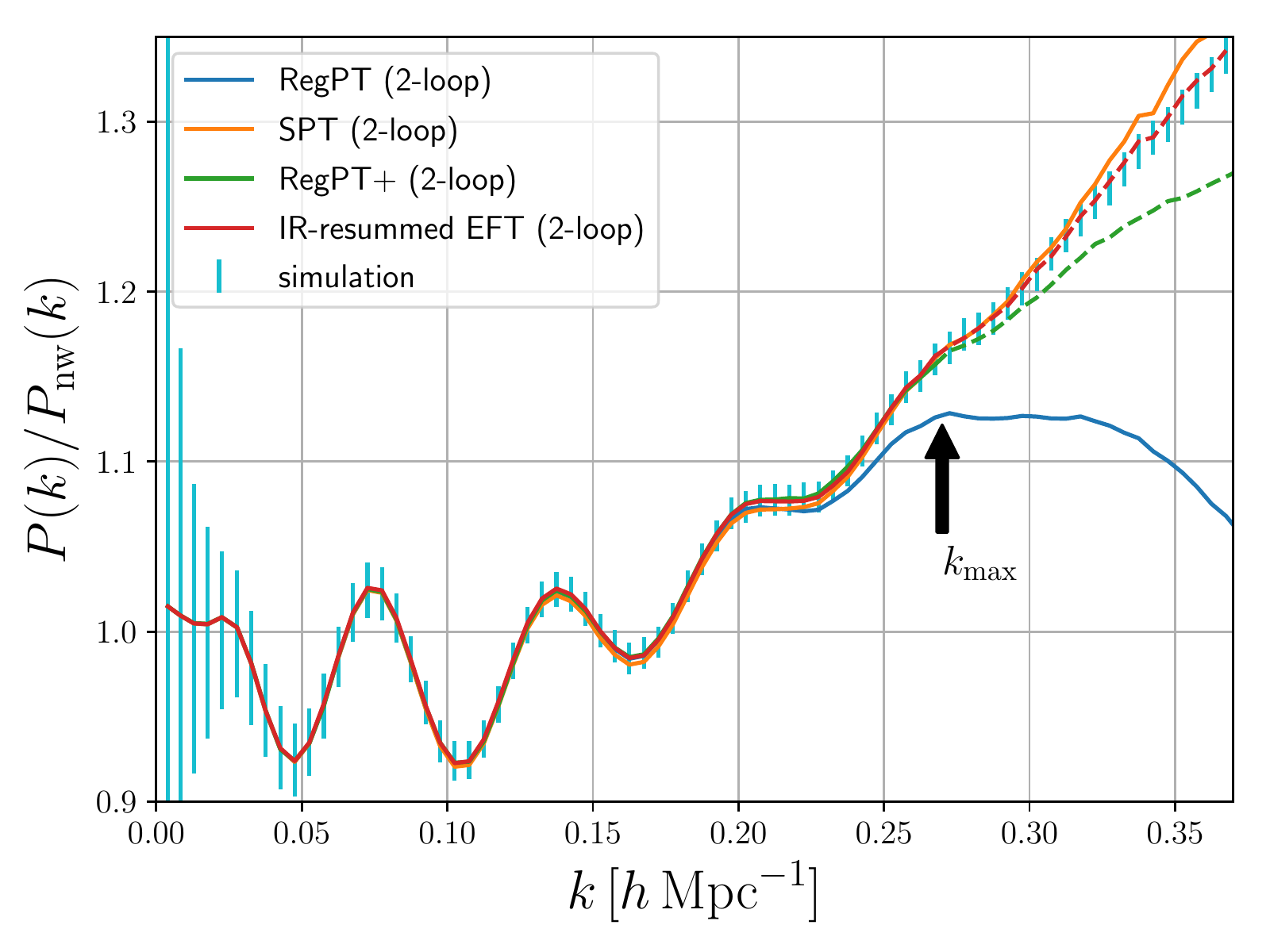}
  \caption{Predictions of power spectrum
  for each analytical method with fiducial cosmological parameters.
  For \texttt{RegPT+} and IR-resummed EFT models,
  nuisance parameters are determined by the least-squares method
  using power spectrum up to $k_\mathrm{max} = 0.27 \, \hMpcinv$.
  The arrow shows the maximum wavenumber in the least-squares method.
  Note that spectra with these two models for wavenumbers larger than $k_\mathrm{max}$
  are shown as dashed lines.}
  \label{fig:fiducial_power}
\end{figure}

\begin{figure}[htbp]
  \centering
  \includegraphics[width=0.45\textwidth]{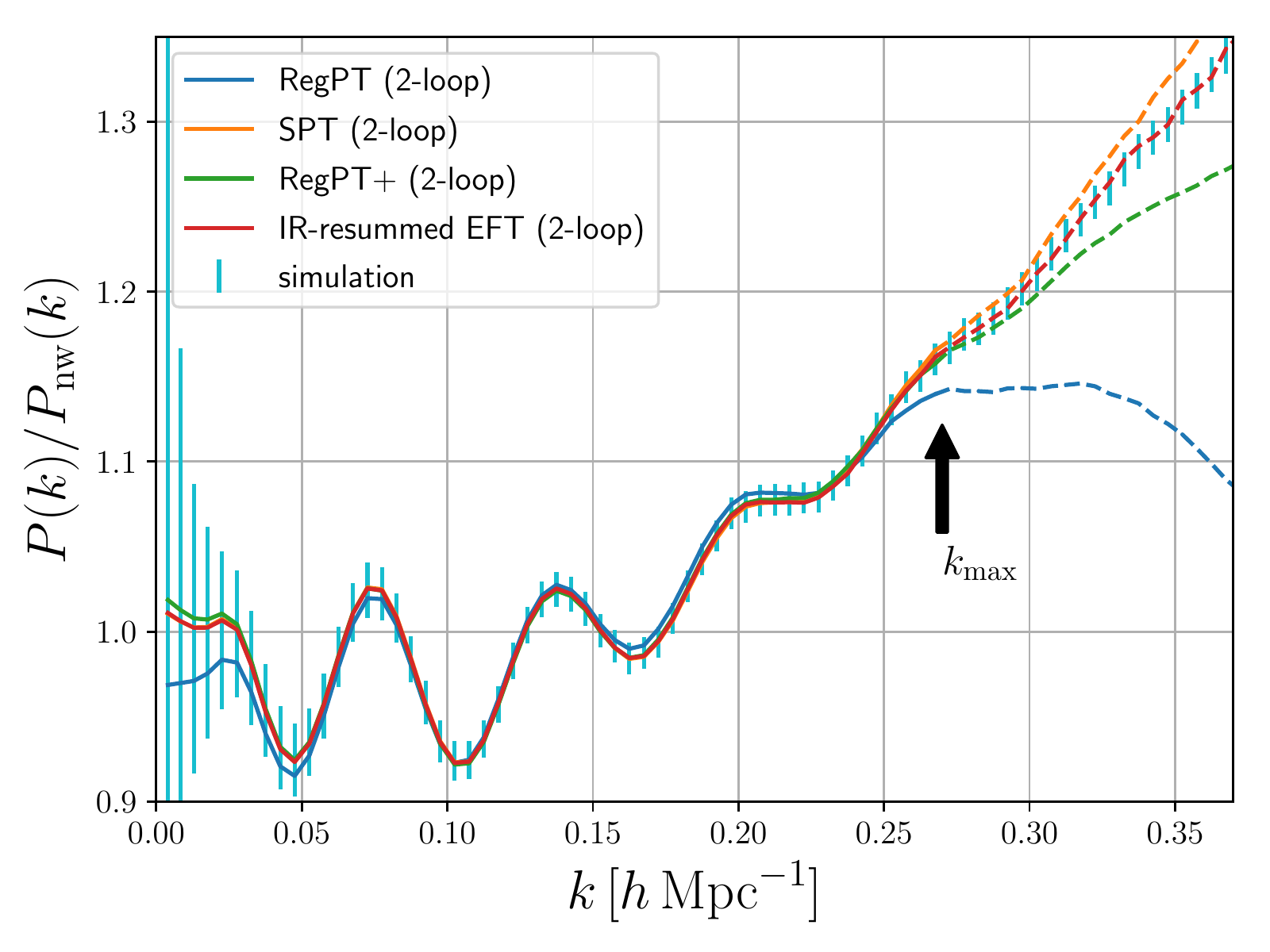}
  \caption{Power spectra with each analytical method
  with best-fit parameters with $k_\mathrm{max} = 0.27 \, \hMpcinv$.
  Note that spectra for wavenumbers larger than $k_\mathrm{max}$
  are shown as dashed lines and the arrow shows $k_\mathrm{max}$.
  The best-fit cosmological parameters and nuisance parameters are estimated
  from MCMC chains.}
  \label{fig:bf_power}
\end{figure}

In Fig.~\ref{fig:fiducial_power}, power spectra computed with input cosmological parameters
and the power spectrum measured from the simulations are shown.
\texttt{RegPT} can accurately fit the power spectrum up to moderate $k \sim 0.2 \, \hMpcinv$,
but for higher $k \gtrsim 0.20 \, \hMpcinv$, it fails to fit the power spectrum.
At a first glance, SPT seems to reproduce the power spectrum at the wide range of scales,
but the small discrepancy can be seen even at large scales $\sim 0.15 \, \hMpcinv$.
In \texttt{RegPT} and SPT, there are no free parameters, but
\texttt{RegPT+} and IR-resummed EFT models contain free parameters,
which are fitted by the least squares method using spectrum up to $0.27 \, \hMpcinv$.
The free parameters help to reproduce the spectrum at small scales,
and the fitting results are improved compared with the cases without free parameters.
In Fig.~\ref{fig:bf_power}, power spectra with best-fit parameters
estimated from MCMC chains with the maximum wavenumber $k_\mathrm{max} = 0.27 \, \hMpcinv$
and the spectrum measured from simulations are shown.
For \texttt{RegPT}, the fitting works well at small scales $\gtrsim 0.20 \, \hMpcinv$,
where errors are small, at the expense of the agreement at large scales.
On the other hand, SPT can reproduce the overall shape of the power spectrum.
Furthermore, with the help of the introduction of free parameters,
\texttt{RegPT+} and IR-resummed EFT can completely capture the feature up to $k_\mathrm{max}$.
However, even if the best-fitting power spectra are consistent,
the best-fit cosmological parameters do not always match with the input parameters.
This aspect is discussed in detail in Sec.~\ref{sec:bias}.
In Figs.~\ref{fig:fiducial_power} and \ref{fig:bf_power},
the results with \texttt{RESPRESSO} are not shown
because this method by construction gives the identical power spectrum measured from simulations.

\begin{figure}[htbp]
  \centering
  \includegraphics[width=0.45\textwidth]{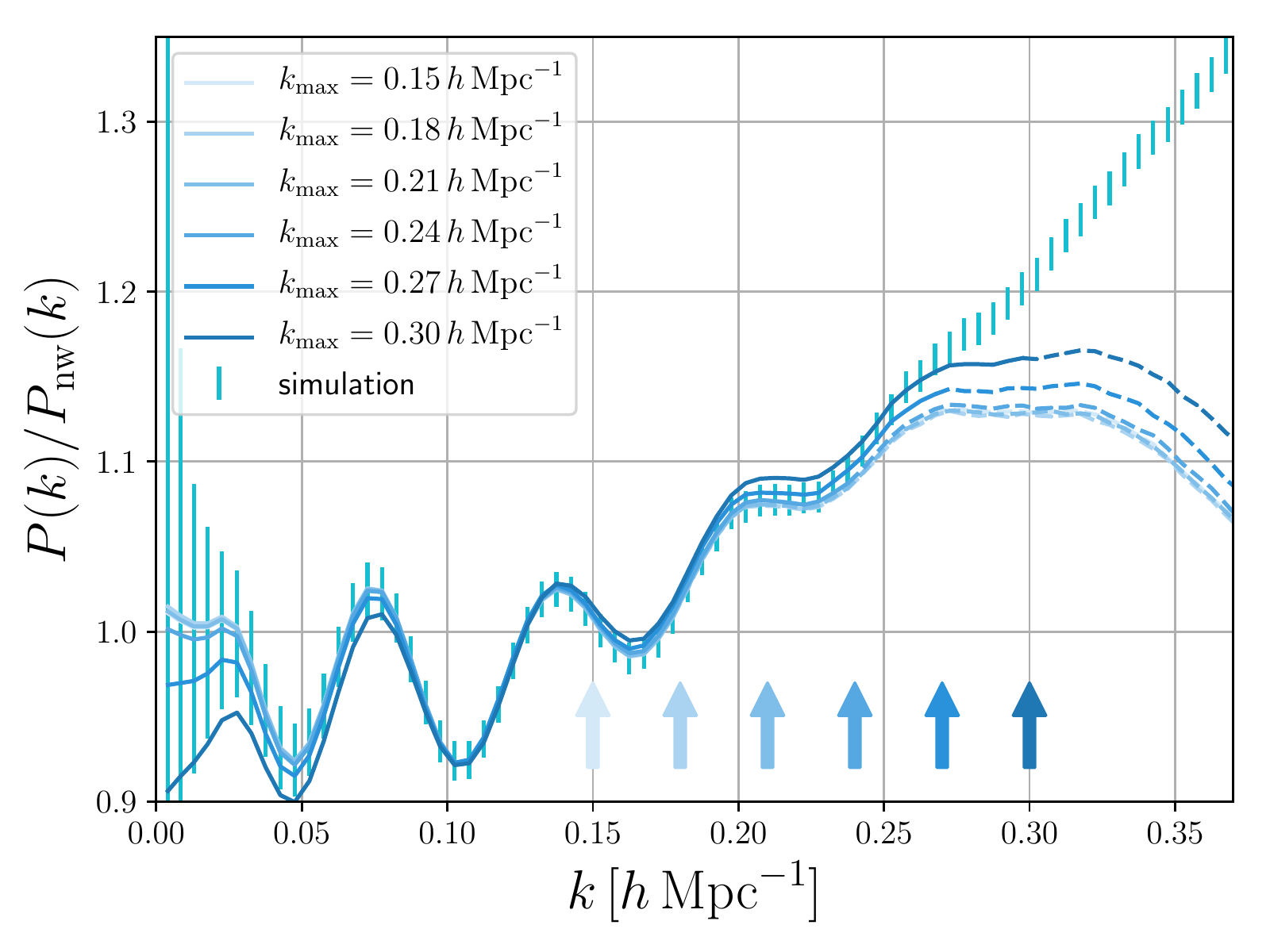}
  \caption{Best-fitting power spectra for \texttt{RegPT} at 2-loop level
  with best-fit parameters with different maximum wavenumbers
  from $0.15 \, \hMpcinv$ to $0.30 \, \hMpcinv$.
  The arrows show corresponding maximum wavenumbers $k_\mathrm{max}$.
  Note that spectra for wavenumbers larger than $k_\mathrm{max}$
  are shown as dashed lines.
  The best-fit cosmological parameters are estimated from MCMC chains.}
  \label{fig:bf_power_2loop}
\end{figure}

\begin{figure}[htbp]
  \centering
  \includegraphics[width=0.45\textwidth]{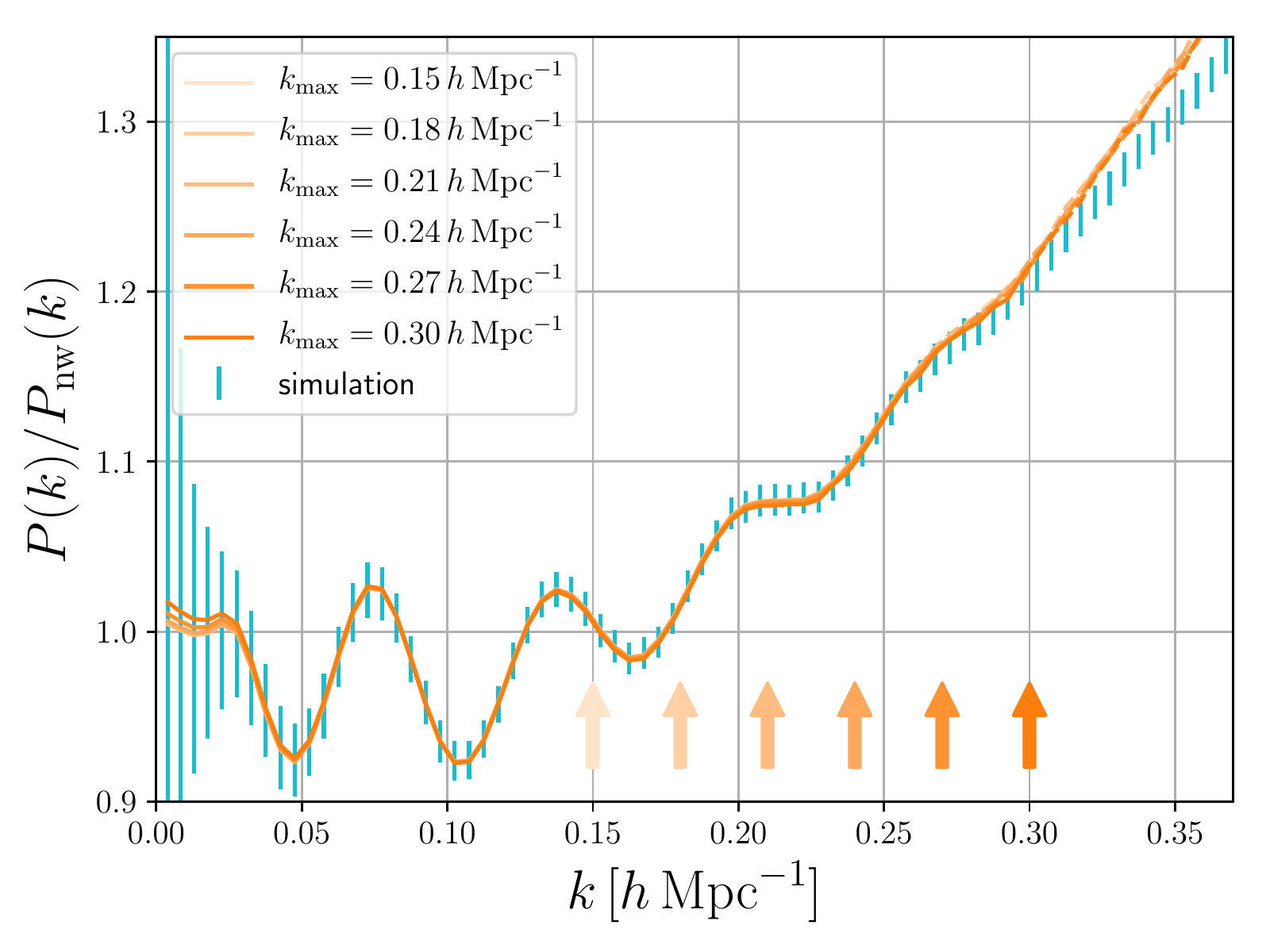}
  \caption{Best-fitting power spectra for SPT at 2-loop level
  with best-fit parameters with different maximum wavenumbers
  from $0.15 \, \hMpcinv$ to $0.30 \, \hMpcinv$.
  The arrows show corresponding maximum wavenumbers $k_\mathrm{max}$.
  Note that spectra for wavenumbers larger than $k_\mathrm{max}$
  are shown as dashed lines.
  The best-fit cosmological parameters are estimated from MCMC chains.}
  \label{fig:bf_power_SPT2loop}
\end{figure}

\begin{figure}[htbp]
  \centering
  \includegraphics[width=0.45\textwidth]{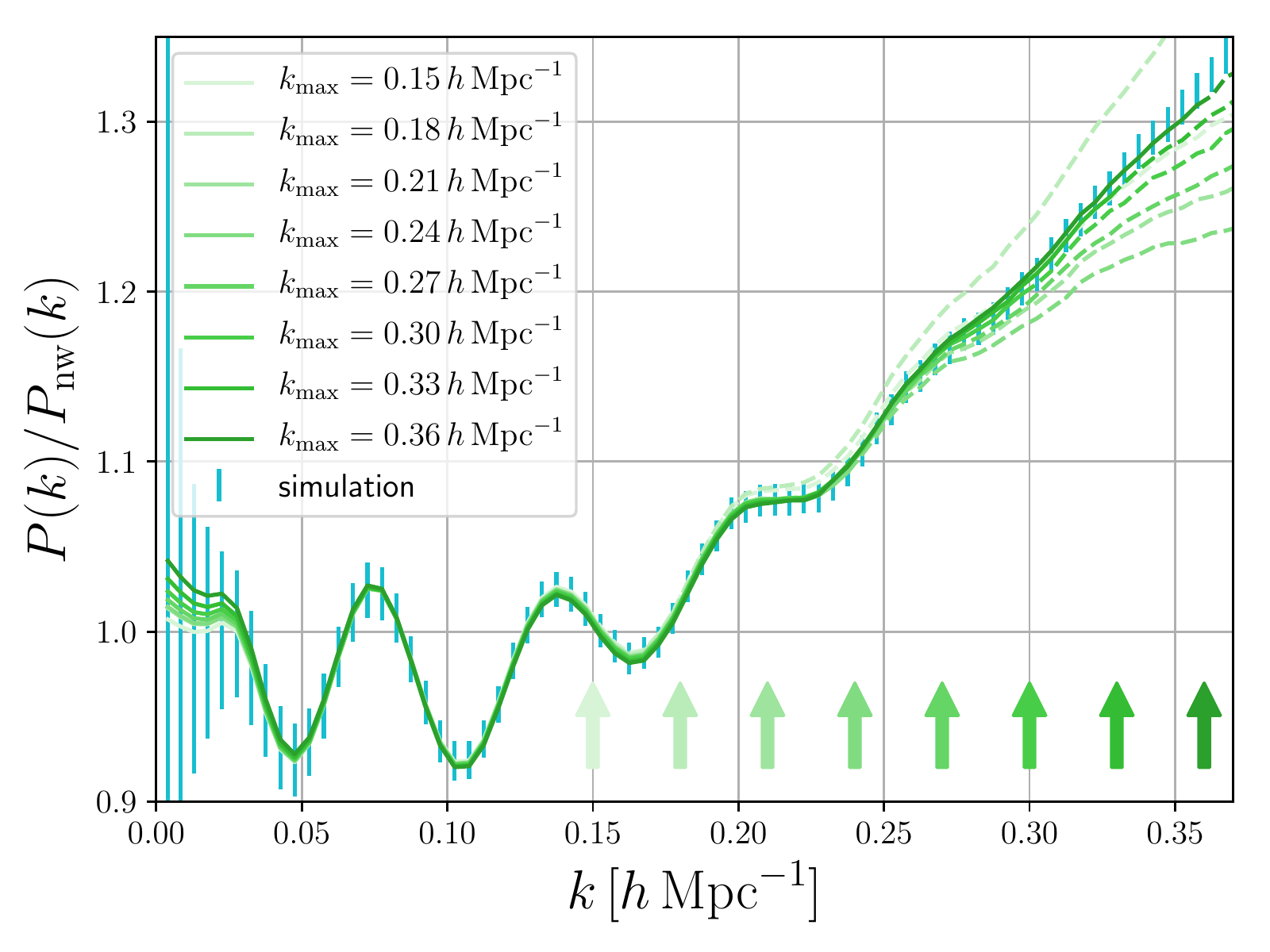}
  \caption{Best-fitting power spectra for \texttt{RegPT+} at 2-loop level
  with best-fit parameters with different maximum wavenumbers
  from $0.15 \, \hMpcinv$ to $0.36 \, \hMpcinv$.
  The arrows show corresponding maximum wavenumbers $k_\mathrm{max}$.
  Note that spectra for wavenumbers larger than $k_\mathrm{max}$
  are shown as dashed lines.
  The best-fit cosmological parameters are estimated from MCMC chains.}
  \label{fig:bf_power_sigma_d}
\end{figure}

\begin{figure}[htbp]
  \centering
  \includegraphics[width=0.45\textwidth]{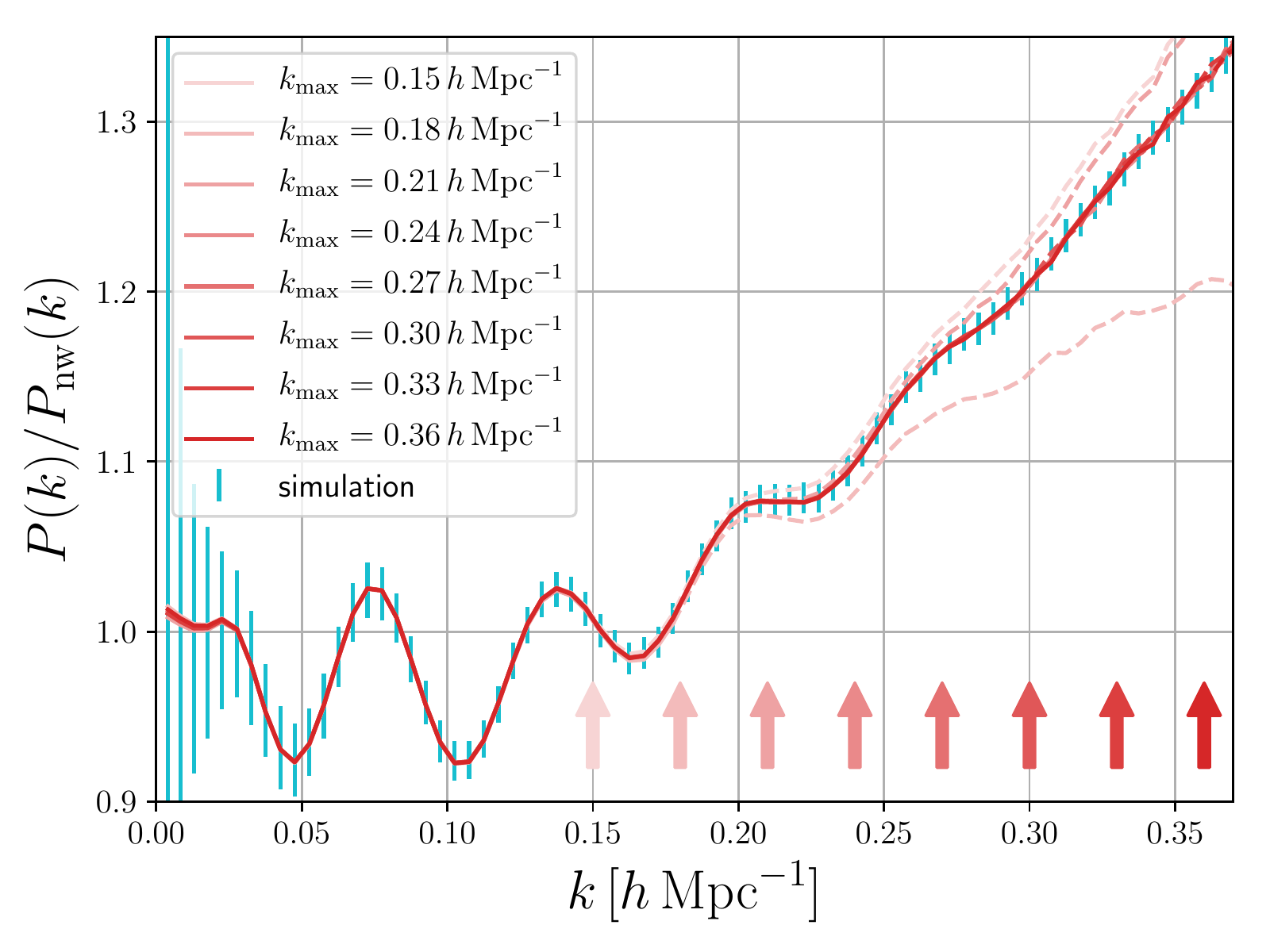}
  \caption{Best-fitting power spectra for IR-resummed EFT at 2-loop level
  with best-fit parameters with different maximum wavenumbers
  from $0.15 \, \hMpcinv$ to $0.36 \, \hMpcinv$.
  The arrows show corresponding maximum wavenumbers $k_\mathrm{max}$.
  Note that spectra for wavenumbers larger than $k_\mathrm{max}$
  are shown as dashed lines.
  The best-fit cosmological parameters are estimated from MCMC chains.}
  \label{fig:bf_power_IRresum}
\end{figure}

We also show results of fitting with different $k_\mathrm{max}$
in Figs.~\ref{fig:bf_power_2loop}, \ref{fig:bf_power_SPT2loop},
\ref{fig:bf_power_sigma_d}, and \ref{fig:bf_power_IRresum}.
In the case of \texttt{RegPT}, for $k_\mathrm{max} \lesssim 0.24 \, \hMpcinv$,
the fitting works well but for larger $k_\mathrm{max}$, it starts to fail
and discrepancy appears at large scales
because the fitting is determined by small-scale spectra where
errors are small but predictions at these scales are no longer reliable,
as we have seen in Fig.~\ref{fig:bf_power}.
SPT can fit the power spectrum well at large scales,
but the small scale feature cannot be captured even with best-fit parameters for large $k_\mathrm{max}$.
As we have seen, \texttt{RegPT+} and IR-resummed EFT can reproduce power spectrum
even for $k_\mathrm{max} > 0.3 \, \hMpcinv$ with the help of the free parameters.

\subsection{Parameter estimation with MCMC analysis}
\begin{figure}[htbp]
  \centering
  \includegraphics[width=0.45\textwidth]{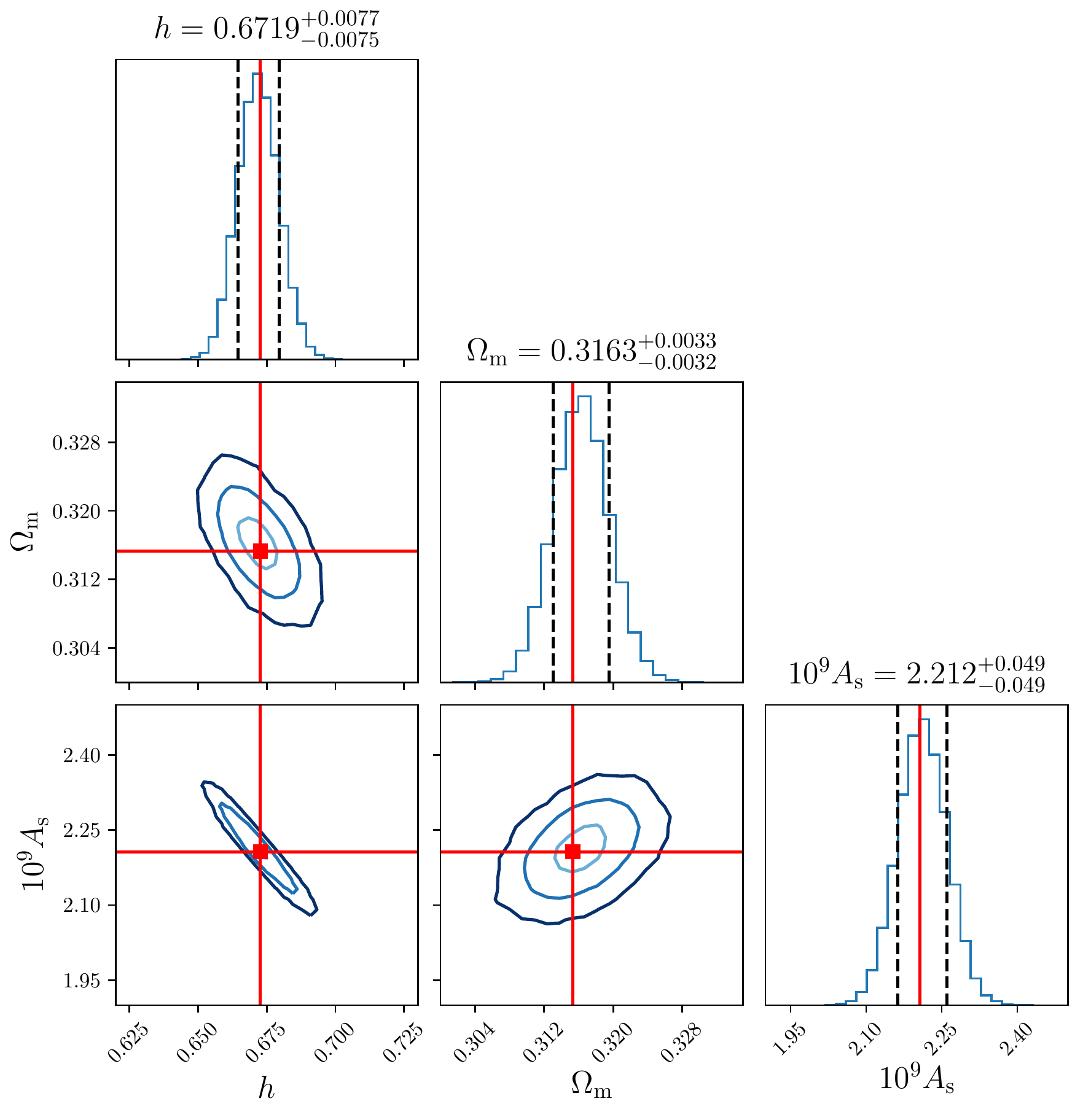}
  \caption{Parameter confidence regions with \texttt{RegPT}
  at 2-loop level and $k_\mathrm{max} = 0.21 \, \hMpcinv$.
  The light, normal, and dark blue lines correspond to the $1\text{-}\sigma$,
  $2\text{-}\sigma$, and $3\text{-}\sigma$ limits, respectively.
  The median and 16\% and 84\% percentiles are
  also shown on the top of each histogram.
  The black dashed lines correspond to 16\% and 84\% percentiles.
  The red lines show the input parameters.}
  \label{fig:confidence_2loop}
\end{figure}

\begin{figure}[htbp]
  \centering
  \includegraphics[width=0.45\textwidth]{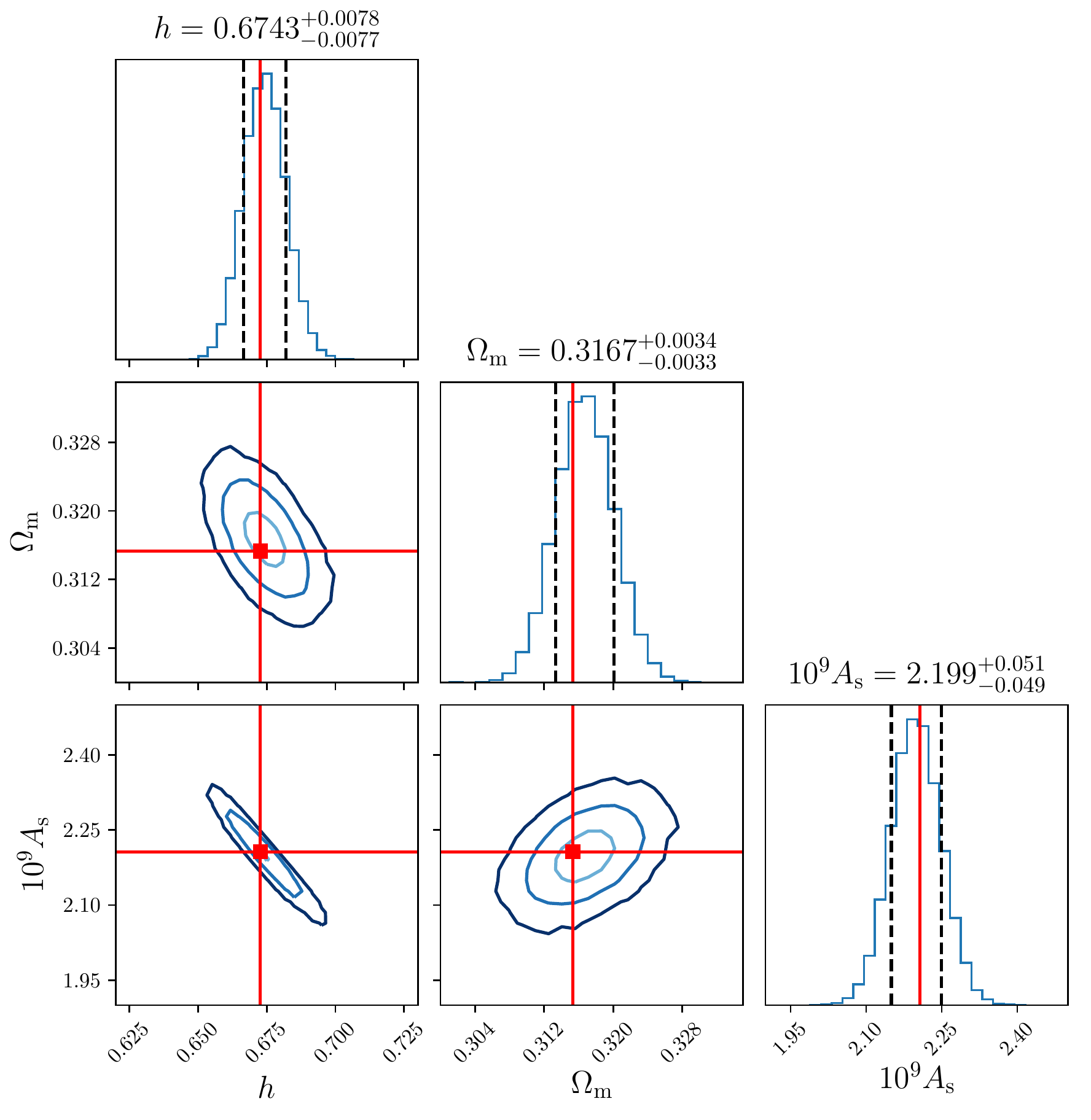}
  \caption{Parameter confidence regions with SPT
  at 2-loop level and $k_\mathrm{max} = 0.21 \, \hMpcinv$.
  The light, normal, and dark blue lines correspond to the $1\text{-}\sigma$,
  $2\text{-}\sigma$, and $3\text{-}\sigma$ limits, respectively.
  The median and 16\% and 84\% percentiles are
  also shown on the top of each histogram.
  The black dashed lines correspond to 16\% and 84\% percentiles.
  The red lines show the input parameters.}
  \label{fig:confidence_SPT2loop}
\end{figure}

\begin{figure}[htbp]
  \centering
  \includegraphics[width=0.45\textwidth]{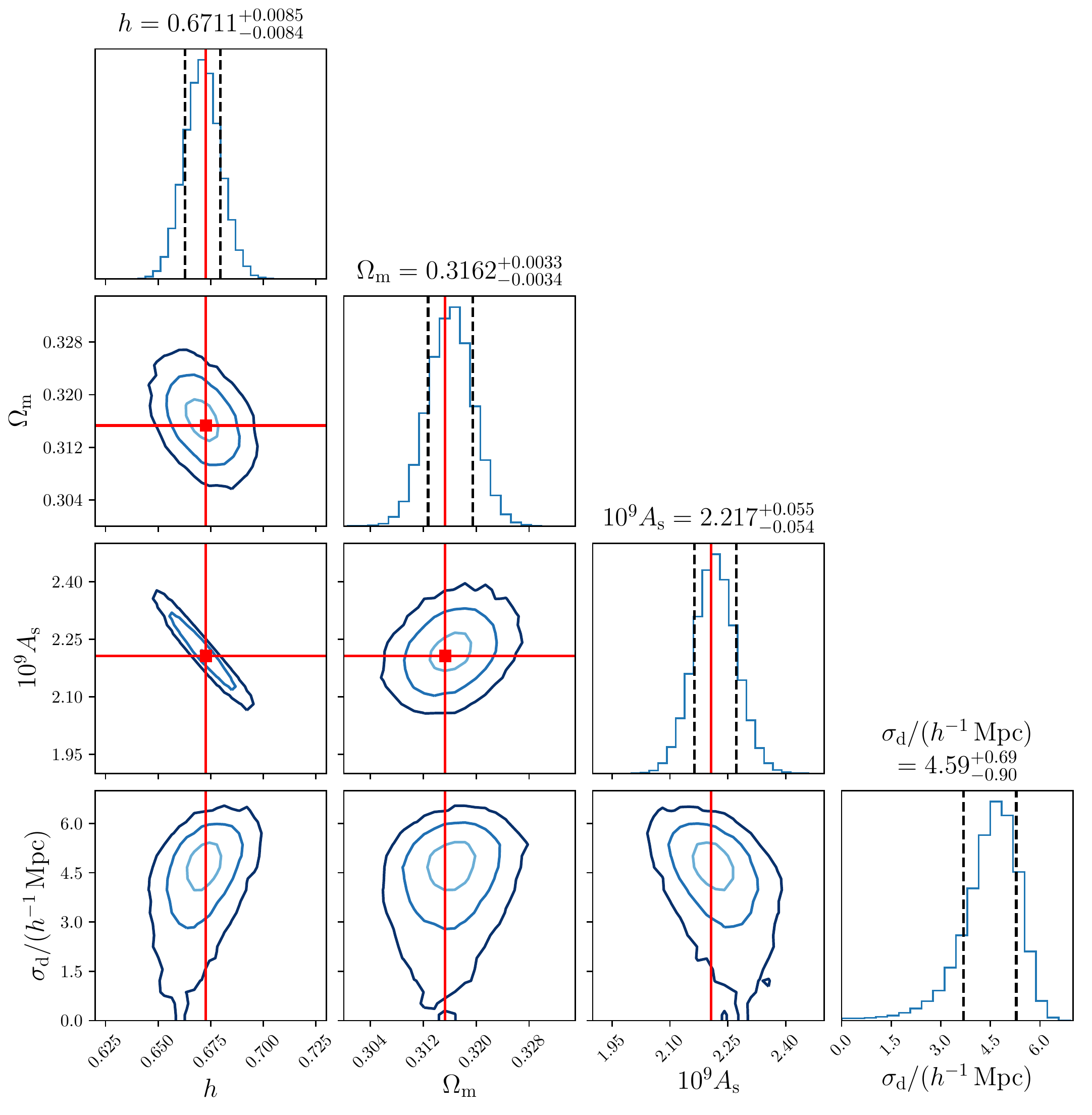}
  \caption{Parameter confidence regions with \texttt{RegPT+}
  at 2-loop level and $k_\mathrm{max} = 0.21 \, \hMpcinv$.
  The light, normal, and dark blue lines correspond to the $1\text{-}\sigma$,
  $2\text{-}\sigma$, and $3\text{-}\sigma$ limits, respectively.
  The median and 16\% and 84\% percentiles are
  also shown on the top of each histogram.
  The black dashed lines correspond to 16\% and 84\% percentiles.
  The red lines show the input parameters.}
  \label{fig:confidence_sigma_d}
\end{figure}

\begin{figure*}[htbp]
  \centering
  \includegraphics[clip,width=0.9\textwidth]{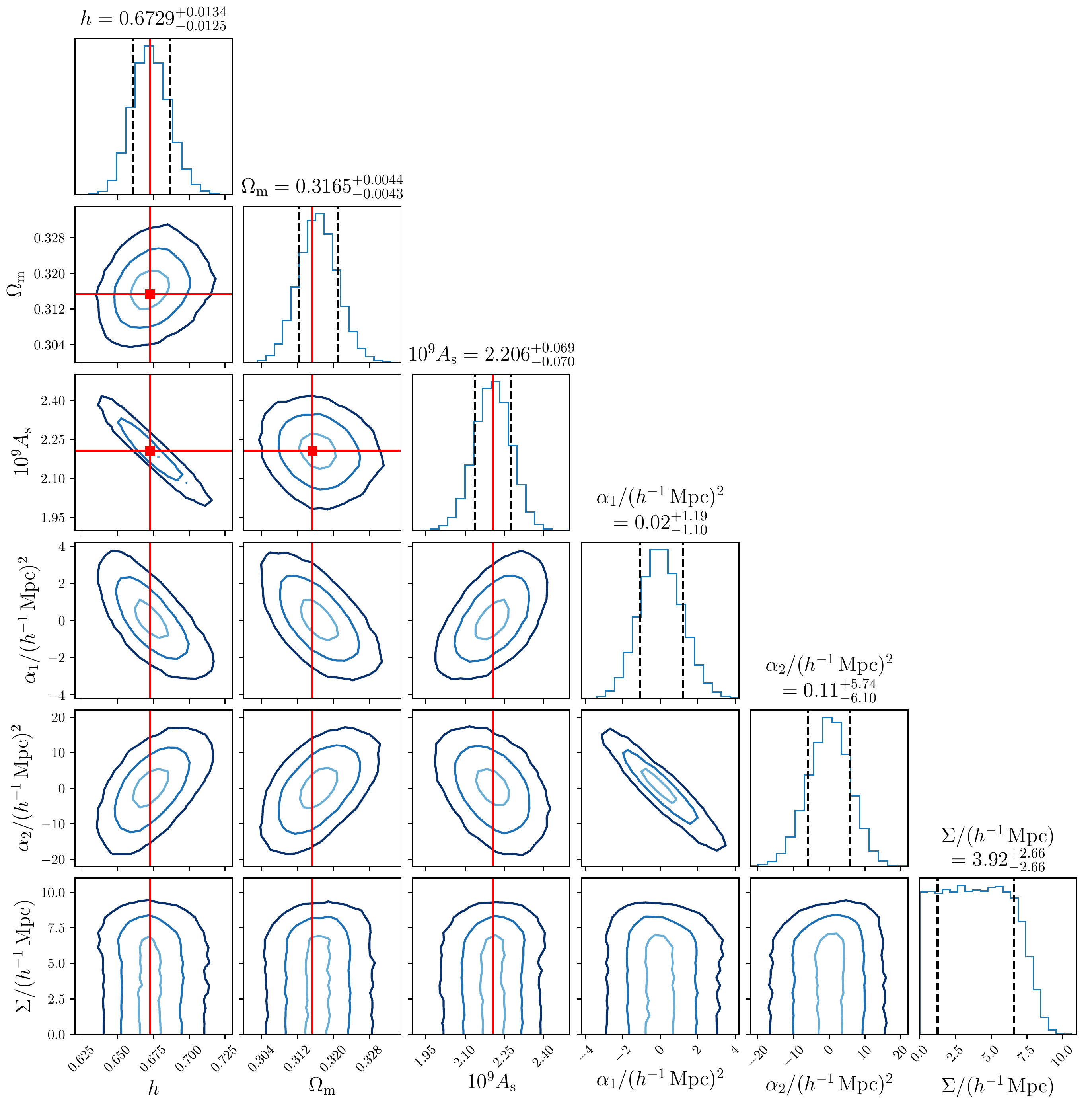}
  \caption{Parameter confidence regions with IR-resummed EFT
  at 2-loop level and $k_\mathrm{max} = 0.21 \, \hMpcinv$.
  The light, normal, and dark blue lines correspond to the $1\text{-}\sigma$,
  $2\text{-}\sigma$, and $3\text{-}\sigma$ limits, respectively.
  The median and 16\% and 84\% percentiles are
  also shown on the top of each histogram.
  The black dashed lines correspond to 16\% and 84\% percentiles.
  The red lines show the input parameters.}
  \label{fig:confidence_IRresum}
\end{figure*}

\begin{figure}[htbp]
  \centering
  \includegraphics[width=0.45\textwidth]{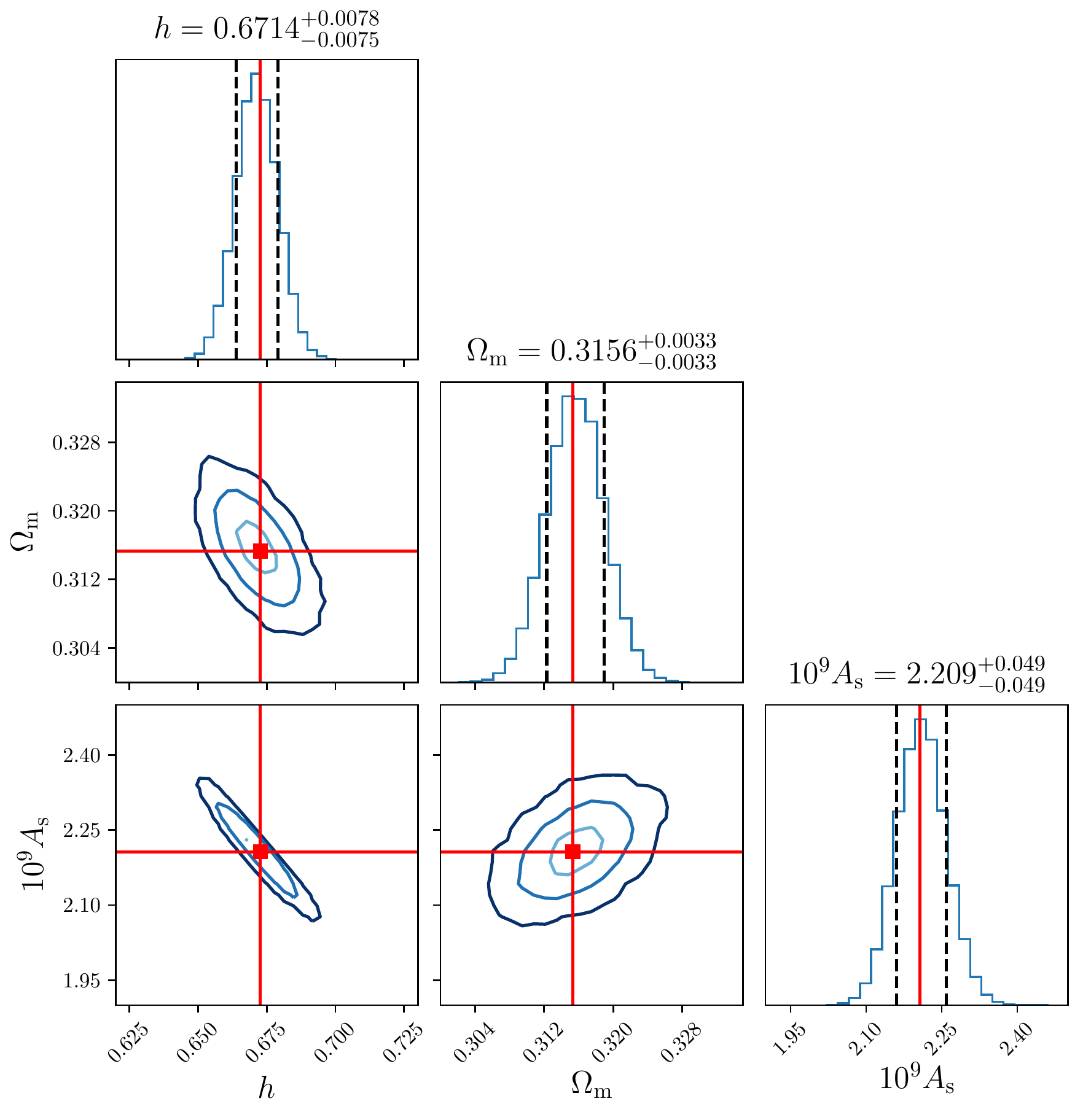}
  \caption{Parameter confidence regions with \texttt{RESPRESSO}
  at 2-loop level and $k_\mathrm{max} = 0.21 \, \hMpcinv$.
  The light, normal, and dark blue lines correspond to the $1\text{-}\sigma$,
  $2\text{-}\sigma$, and $3\text{-}\sigma$ limits, respectively.
  The median and 16\% and 84\% percentiles are
  also shown on the top of each histogram.
  The black dashed lines correspond to 16\% and 84\% percentiles.
  The red lines show the input parameters.}
  \label{fig:confidence_RESPRESSO}
\end{figure}

In Figs.~\ref{fig:confidence_2loop}, \ref{fig:confidence_SPT2loop},
\ref{fig:confidence_sigma_d}, \ref{fig:confidence_IRresum},
and \ref{fig:confidence_RESPRESSO},
the confidence regions of cosmological parameters
with \texttt{RegPT}, SPT, \texttt{RegPT+}, and IR-resummed EFT at 2-loop level
and \texttt{RESPRESSO} for $k_\mathrm{max} = 0.21 \, \hMpcinv$ are shown.
At this scale, all prescriptions give the precise predictions
of power spectrum and we can safely recover the input cosmological parameters.
However, for example, the constraints with \texttt{RegPT} are
tighter than those with IR-resummed EFT
because free parameters introduced in this model degrade the resultant constraints.
In addition, these parameters also affect parameter degeneracy
between cosmological parameters.
This effect will be addressed later in Sec.~\ref{sec:correlations}.
For other models which contain no nuisance parameters, i.e.,
SPT and \texttt{RESPRESSO}, the constraints are almost the same as those with \texttt{RegPT}.
On the other hand, for \texttt{RegPT+}, the constraints are weaker than those with \texttt{RegPT}
as this model also introduces a free parameter.

\begin{figure*}[htbp]
  \centering
  \includegraphics[width=0.9\textwidth]{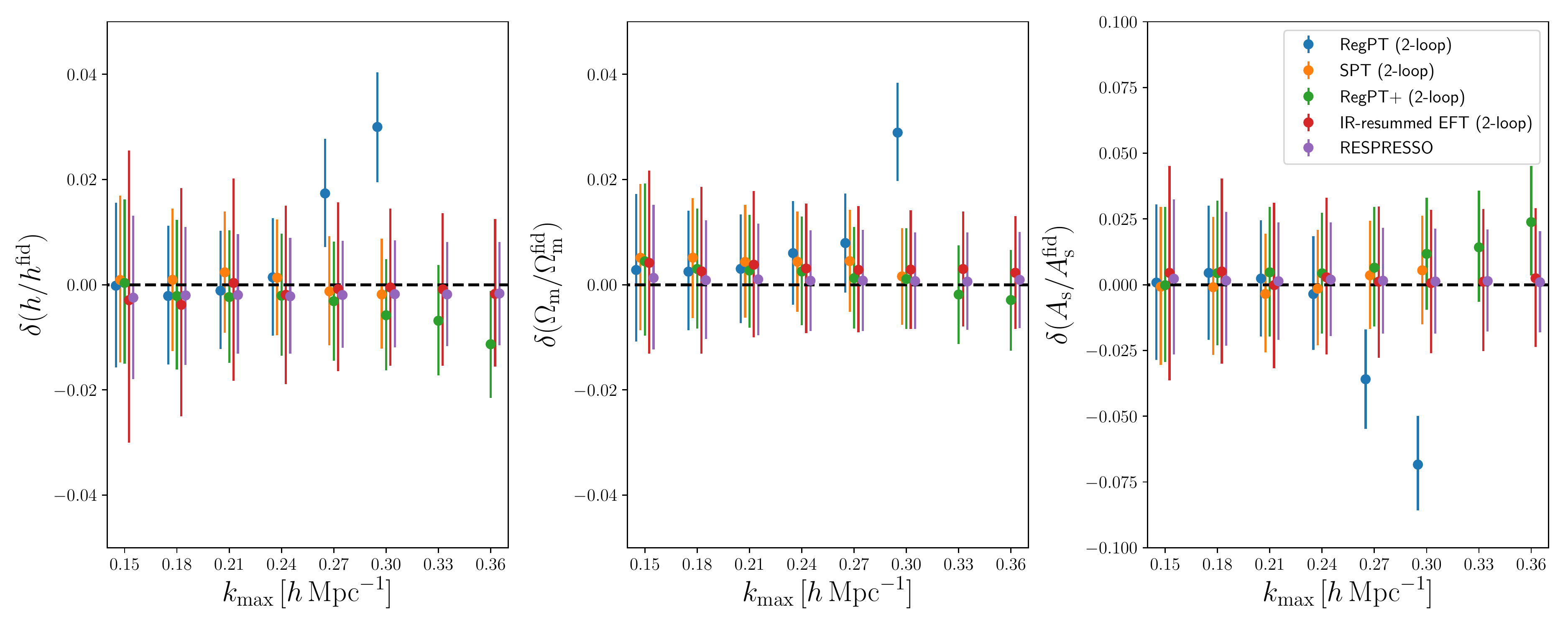}
  \caption{Medians of cosmological parameters estimated from MCMC chains
  with errors, shown as the fractional ratios with respect to the fiducial values.
  The lower (upper) limit of error bars
  correspond to $16 \%$ ($84 \%$) percentile.}
  \label{fig:param}
\end{figure*}

\begin{figure*}[htbp]
  \centering
  \includegraphics[width=0.9\textwidth]{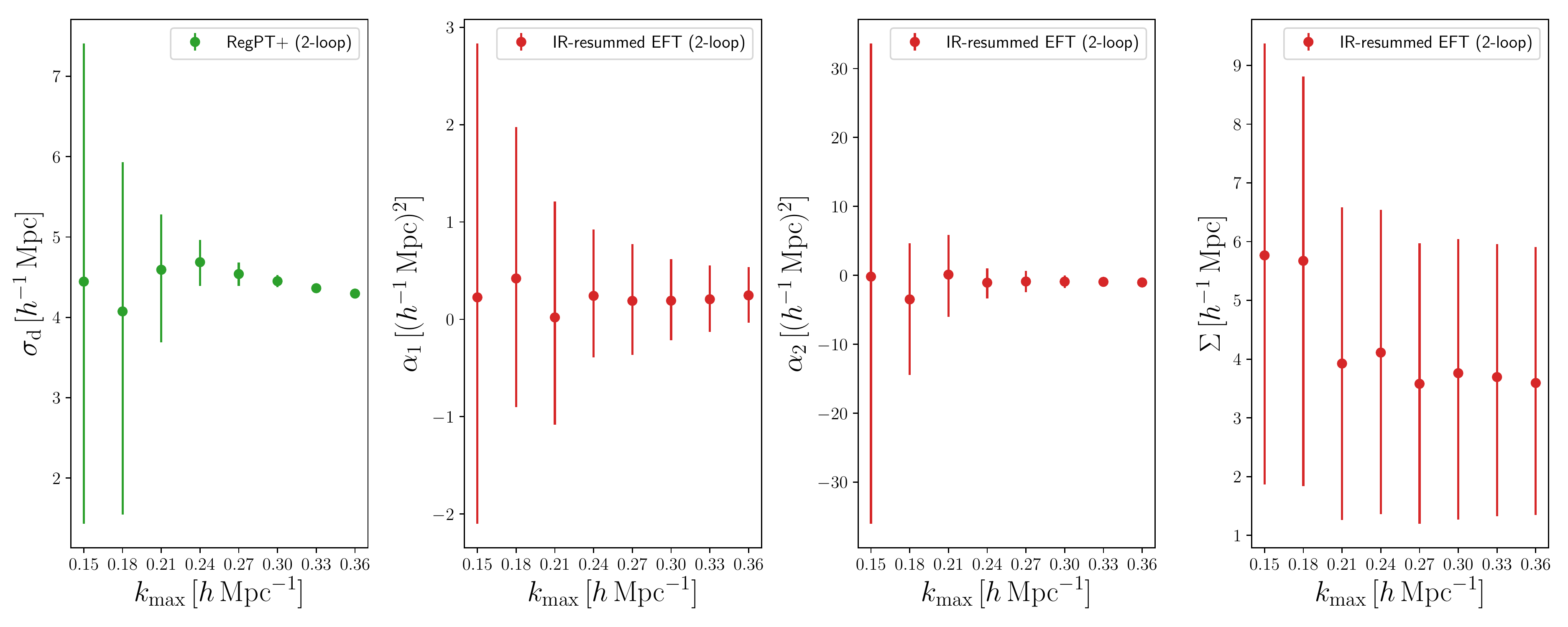}
  \caption{Medians of nuisance parameters,
  $\sigmad$ for \texttt{RegPT+}
  and $\alpha_1$, $\alpha_2$, and $\Sigma$ for IR-resummed EFT,
  estimated from MCMC chains.
  The lower (upper) limit of error bars
  correspond to $16 \%$ ($84 \%$) percentile.}
  \label{fig:nuisance}
\end{figure*}

Next, we show the estimated cosmological and nuisance parameters
with analytical models at 2-loop level in Figs.~\ref{fig:param} and \ref{fig:nuisance}.
The values plotted in these figures are medians, which are robust to outliers,
instead of means. However, for cosmological parameters,
since the posterior distributions are symmetric, the median and mean are almost the same.
On the other hand, for some of nuisance parameters
(see, e.g., the parameter $\Sigma$ in Fig.~\ref{fig:confidence_IRresum}),
the posterior distribution is far from symmetric,
the median can be different from the sample mean.
The range of error bars corresponds to 16\% and 84\% percentiles,
which correspond to $1\text{-}\sigma$ range
when the posterior distribution is completely Gaussian.
For \texttt{RegPT} at 2-loop level,
this model gives unbiased estimates up to $k_\mathrm{max} \lesssim 0.24 \, \hMpcinv$,
where the calculations are supposed to be accurate.
As a general trend, the errors become small by increasing $k_\mathrm{max}$
because more information become available.
However, at high $k \gtrsim 0.25 \, \hMpcinv$,
this model is not supposed to compute the spectrum accurately
(see Fig.~\ref{fig:fiducial_power}).
As a result, the fitting process itself breaks down, and then errors increase and
estimated parameters deviate from the true values.
\texttt{RegPT+}, IR-resummed EFT, and \texttt{RESPRESSO}
can all reproduce the input cosmological parameters up to high $k_\mathrm{max}$.
However, \texttt{RegPT+} and IR-resummed EFT
contain free nuisance parameters and thus they have more degrees of freedom to
fit the power spectrum. Thus, that leads to degradation of constraints.

\subsection{Figure of bias}
\label{sec:bias}
\begin{figure}[htbp]
  \centering
  \includegraphics[width=0.45\textwidth]{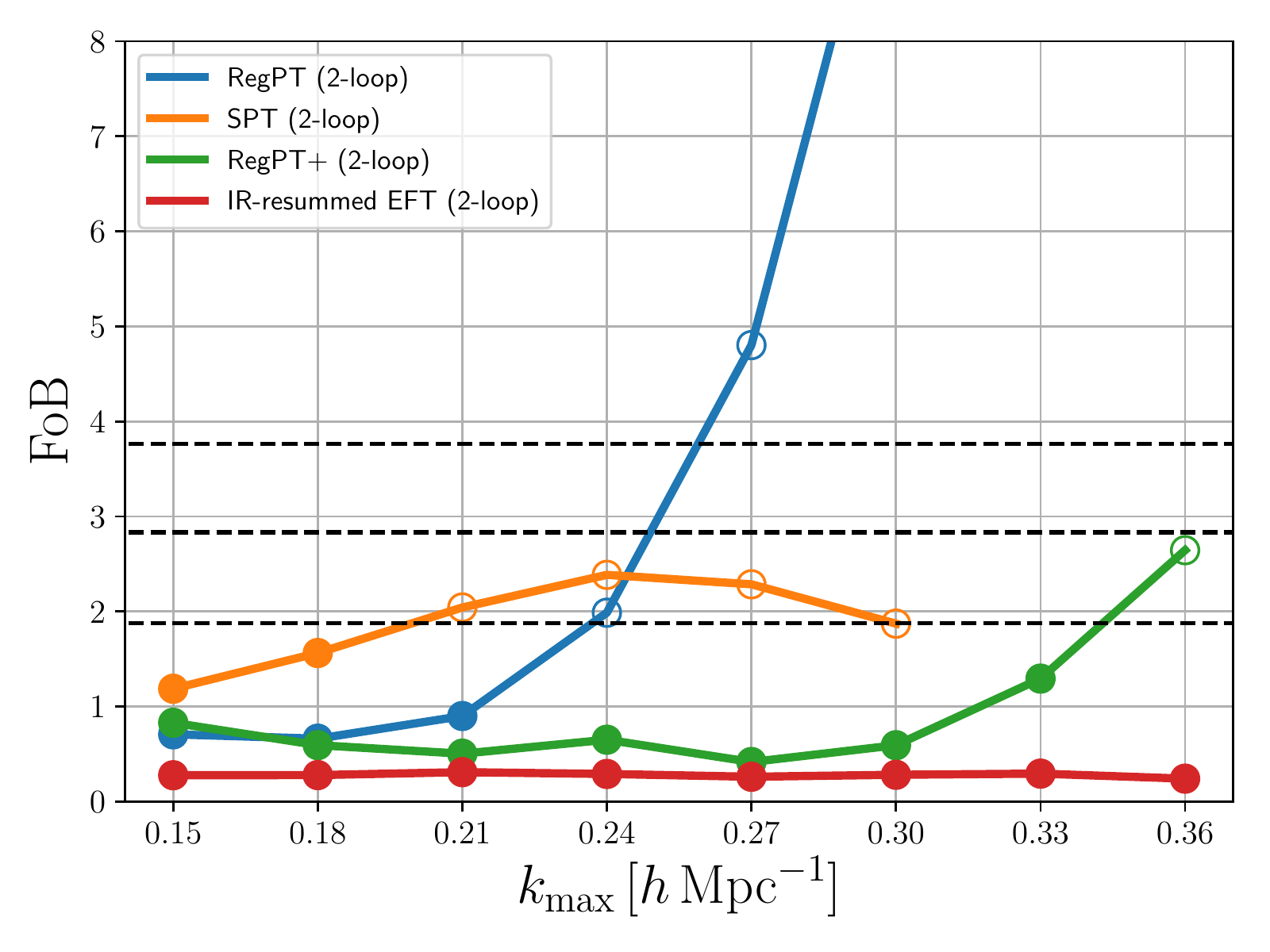}
  \caption{Figure of bias for different models estimated from MCMC chains.
  The black dashed lines show the $1\text{-}\sigma$, $2\text{-}\sigma$, and
  $3\text{-}\sigma$ critical values $1.88$, $2.83$, and $3.76$, respectively.
  The open (filled) symbols represent that the figure of bias
  exceeds (falls below) the $1\text{-}\sigma$ critical value.}
  \label{fig:bias}
\end{figure}

Here, we quantify how close to the input cosmological parameters
estimated parameters are for each template with respect to the statistical errors.
For this purpose, first we compute the correlation matrix
of all parameters $S$, i.e., cosmological parameters
and nuisance parameters ($\sigmad$ for \texttt{RegPT+}
and $\alpha_1$, $\alpha_2$, and $\Sigma$ for IR-resummed EFT),
which can be estimated from MCMC chains:
\beq
S_{\alpha \beta} = \frac{1}{N-1} \sum_k^N (\theta_\alpha^k - \bar{\theta}_\alpha)
(\theta_\beta^k - \bar{\theta}_\beta) ,
\eeq
where $\bm{\theta}^k$ is a parameter vector at the $k$-th step,
$N$ is the number of total steps in chains,
and $\bar{\bm{\theta}}$ is the sample mean of the parameters.
We are interested only in cosmological parameters and marginalize
the posterior distribution over the nuisance parameters.
Under the assumption that the parameters follow the multivariate Gaussian
distribution, this operation simply corresponds to
taking submatrix, which is denoted by $\tilde{S}$.
Then, we define the figure of bias (FoB) as,
\beq
\mathrm{FoB} \equiv \left[ \sum_{\alpha, \beta} \delta \theta_\alpha
\left( \tilde{S} \right)^{-1}_{\alpha \beta}
\delta \theta_\beta \right]^{1/2} ,
\eeq
where $\delta \bm{\theta}$ is the difference between the estimated and input
cosmological parameters \cite{Taruya11}.
FoB corresponds to the distance between true and estimated
cosmological parameters normalized by their variances.
In Fig.~\ref{fig:bias}, we show FoBs with different models and $k_\mathrm{max}$
along with $1\text{-}\sigma$, $2\text{-}\sigma$, and
$3\text{-}\sigma$ critical values, which correspond to $68\%$, $95\%$, and $99.7\%$
percentiles, respectively, in the case of three parameters,
when the parameter deviation $\delta \bm{\theta}$ follows multivariate Gaussian.
These critical values are derived from cumulative distribution function for
the chi-squared distribution. The degree of freedom is 3
regardless of the choice of the model because
the nuisance parameters have already been marginalized.
FoBs of SPT and \texttt{RegPT} exceed the $1\text{-}\sigma$ critical value
from relatively small $k_\mathrm{max}$
while FoBs of \texttt{RegPT+} and IR-resummed EFT
are kept small even for high $k_\mathrm{max}$.
This fact means that we can employ \texttt{RegPT+} and IR-resummed EFT
up to scales $\gtrsim 0.30 \, \hMpcinv$
without having a substantial biased parameter estimation.
However, there is a caveat about the small FoBs.
Since FoB is normalized by the variance of parameters,
if power of constraining parameters is weak, the resultant FoB will be also small.
In the cases of \texttt{RegPT+} and IR-resummed EFT,
the nuisance parameters degrade constraints and that leads to large variances.
Though these models can provide us with the accurate prediction even at small scales,
their small FoBs should be taken with cautions.
The FoB of \texttt{RESPRESSO} is not shown in Fig.~\ref{fig:bias}
because, as stated before, the simulations used in the analysis is also used to calibrate
the response function in \texttt{RESPRESSO} and thus the
FoB of \texttt{RESPRESSO} should always be zero.

So far, we have presented FoBs with power spectra at 2-loop level,
but as a whole, those with spectra at 1-loop level show
similar behavior qualitatively.
However, FoBs are generally larger and exceed $1\text{-}\sigma$ limit
even for smaller $k_\mathrm{max}$.
We present detailed discussions for results at 1-loop level
in Appendix~\ref{sec:1loop_results}.

\subsection{Figure of merit}
\label{sec:merit}
Next, we quantify the precision of parameter estimation for each model
using figure of merit (FoM).
We define FoM as the inverse of volume of the parameter space
determined by iso-posterior density surface \cite{Albrecht06}, i.e.,
\beq
\mathrm{FoM} \equiv \frac{1}{\sqrt{\mathrm{det} \tilde{S} }} .
\label{eq:FoM}
\eeq
Roughly speaking, FoM is an indicator of constraining power for each model.
We also introduce an analytical way to estimate the upper limit of FoM
for \texttt{RESPRESSO}.
With response function approach,
we can compute the Fisher information matrix.
Since the covariance matrix does not depend on parameters,
the Fisher matrix can be given as
\beq
F_{\alpha \beta} = \sum_{k_i, k_j < k_\mathrm{max}}
\frac{\partial P (k_i)}{\partial \theta_\alpha}
(C^{-1})_{ij} \frac{\partial P (k_j)}{\partial \theta_\beta} .
\eeq
The derivatives of the power spectrum
can be obtained via the response function (Eq.~\ref{eq:response}),
\beq
\frac{\partial P (k)}{\partial \theta_\alpha} =
\int dq \frac{\delta P (k)}{\delta P_\Lrm (q)}
\frac{\partial P_\Lrm (q)}{\partial \theta_\alpha} =
\int d \ln q \, K(k, q) \frac{\partial P_\Lrm (q)}{\partial \theta_\alpha} .
\eeq
Then the Fisher matrix estimate of the FoM is given by,
\beq
\mathrm{FoM}_F = \sqrt{\det F} .
\eeq
From the Cram\'er-Rao bound \cite{Albrecht09},
FoMs for \texttt{RESPRESSO} can not exceed $\mathrm{FoM}_F$.
If the likelihood distribution (Eq.~\ref{eq:posterior})
follows Gaussian with respect to parameters,
$\mathrm{FoM}_F$ coincides with
FoM computed from correlation matrix (Eq.~\ref{eq:FoM}).

\begin{figure}[htbp]
  \centering
  \includegraphics[width=0.45\textwidth]{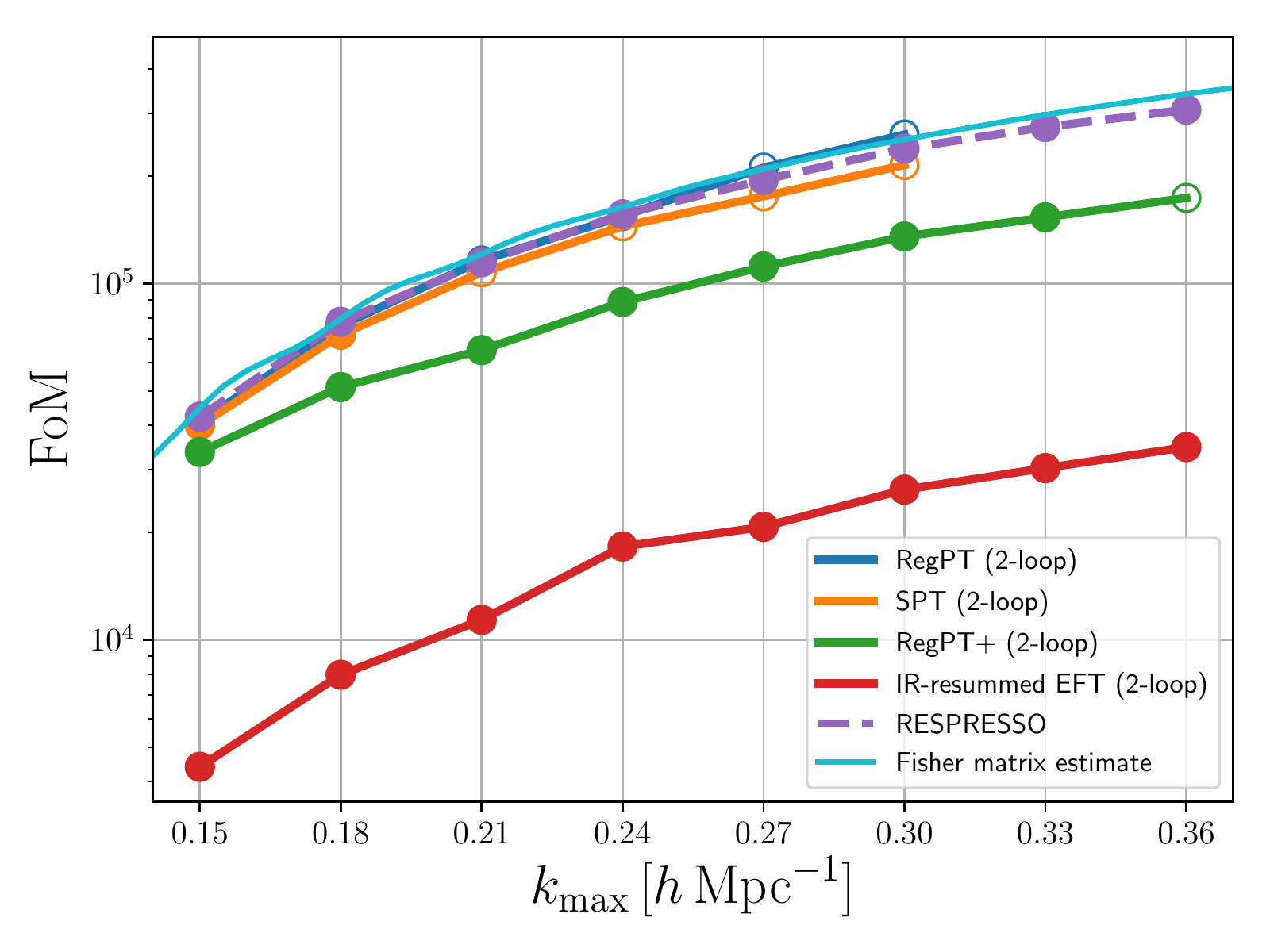}
  \caption{Figure of merit for different models estimated from MCMC chains.
  The result with \texttt{RESPRESSO} is shown as a dashed line
  since it is different from other methods in the sense that
  it relies on simulation-aided approach, not completely analytical prescription.
  The cyan line shows figure of merit from Fisher matrix with \texttt{RESPRESSO}.
  The open (filled) symbols represent that the corresponding figure of bias
  exceeds (falls below) the $1\text{-}\sigma$ critical value (see Fig.~\ref{fig:bias}).}
  \label{fig:merit}
\end{figure}

In Fig.~\ref{fig:merit}, we show FoMs with different models and $k_\mathrm{max}$
and Fisher matrix approach with $\texttt{RESPRESSO}$.
Generally, FoM increases with $k_\mathrm{max}$
because more information becomes available.
As a whole, \texttt{RegPT}, SPT, and \texttt{RESPRESSO} work quite well.
However, shown in Fig.~\ref{fig:bias},
FoBs of \texttt{RegPT} and SPT soon exceed the $1\text{-}\sigma$ limit
leading to biased estimated parameters.
On the other hand, FoMs of \texttt{RegPT+} and IR-resummed EFT
are significantly suppressed with respect to \texttt{RESPRESSO} results.
Clearly though free parameters contained in these models help to
fit the power spectra down to small scales,
the presence of extra degrees of freedom degrades
the constraining power of these models.

\subsection{Correlations between parameters}
\label{sec:correlations}
\begin{figure}[htbp]
  \centering
  \includegraphics[width=0.45\textwidth]{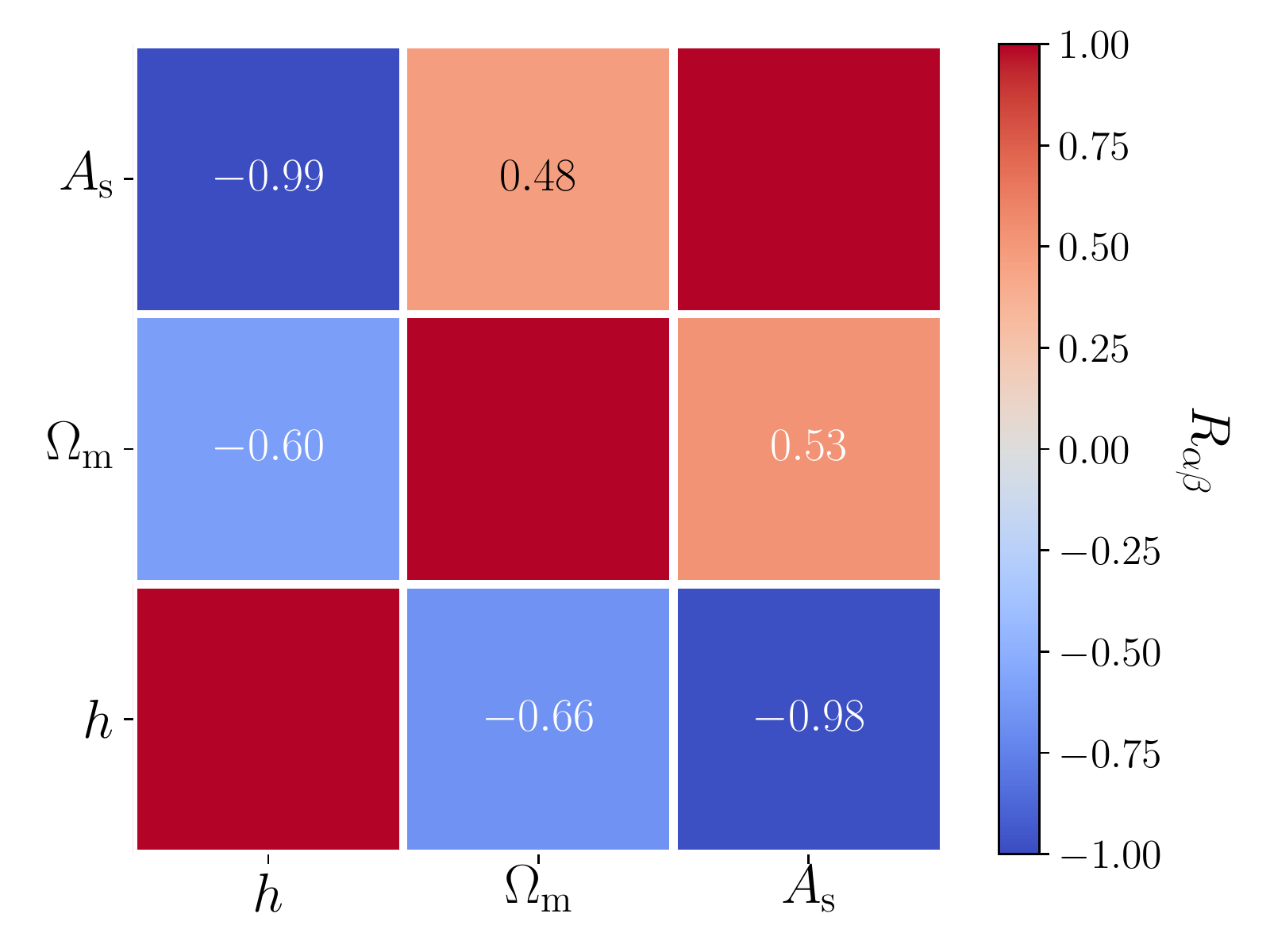}
  \caption{Correlation coefficients for \texttt{RegPT} at 2-loop level.
  The upper left (lower right) triangle shows results
  with $k_\mathrm{max} = 0.18 \, (0.24) \, \hMpcinv$.
  The red (blue) parts correspond to positive (negative) correlations.}
  \label{fig:ccoef_2loop}
\end{figure}

\begin{figure}[htbp]
  \centering
  \includegraphics[width=0.45\textwidth]{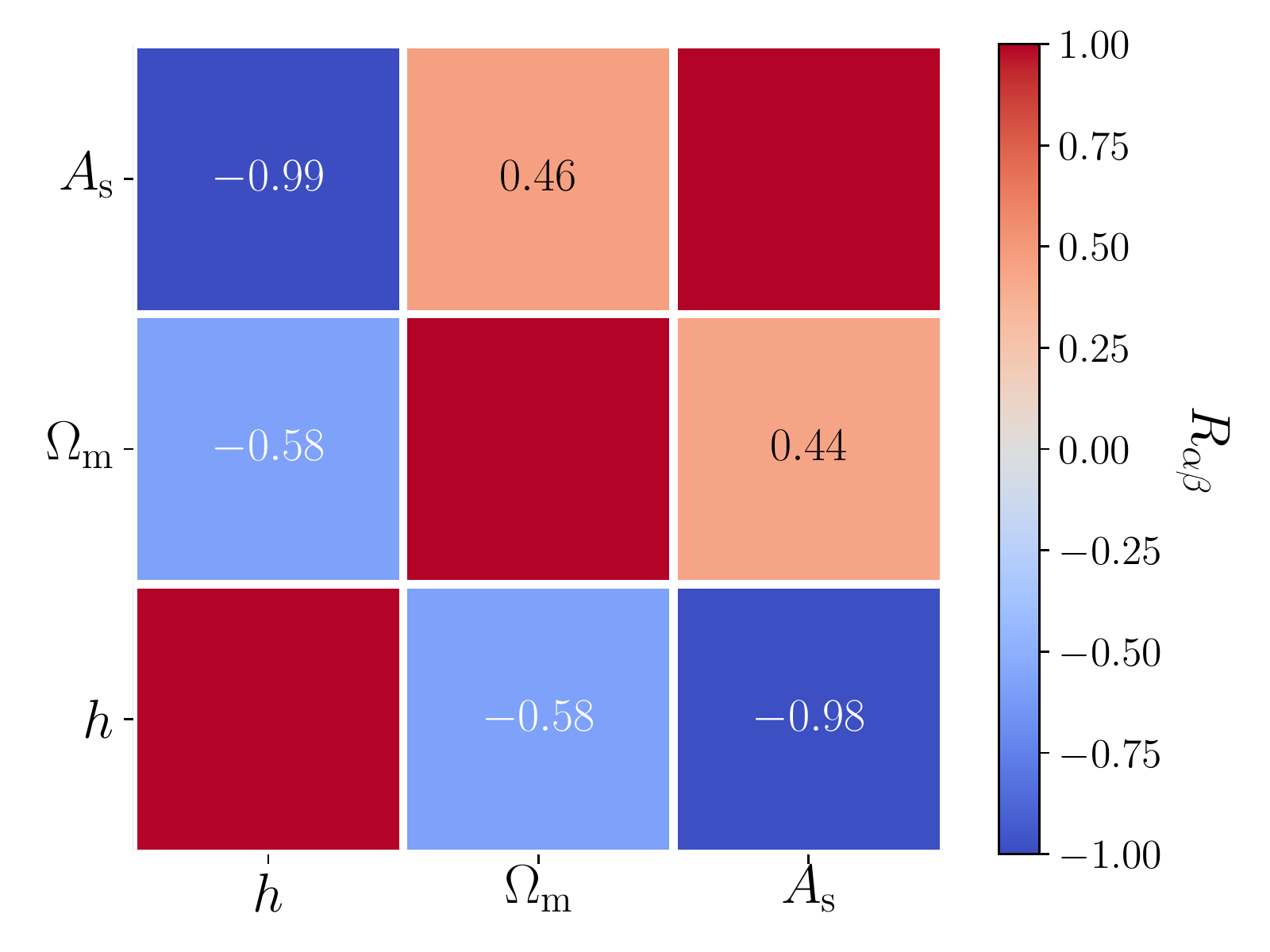}
  \caption{Correlation coefficients for SPT at 2-loop level.
  The upper left (lower right) triangle shows results
  with $k_\mathrm{max} = 0.18 \, (0.24) \, \hMpcinv$.
  The red (blue) parts correspond to positive (negative) correlations.}
  \label{fig:ccoef_SPT2loop}
\end{figure}

\begin{figure}[htbp]
  \centering
  \includegraphics[width=0.45\textwidth]{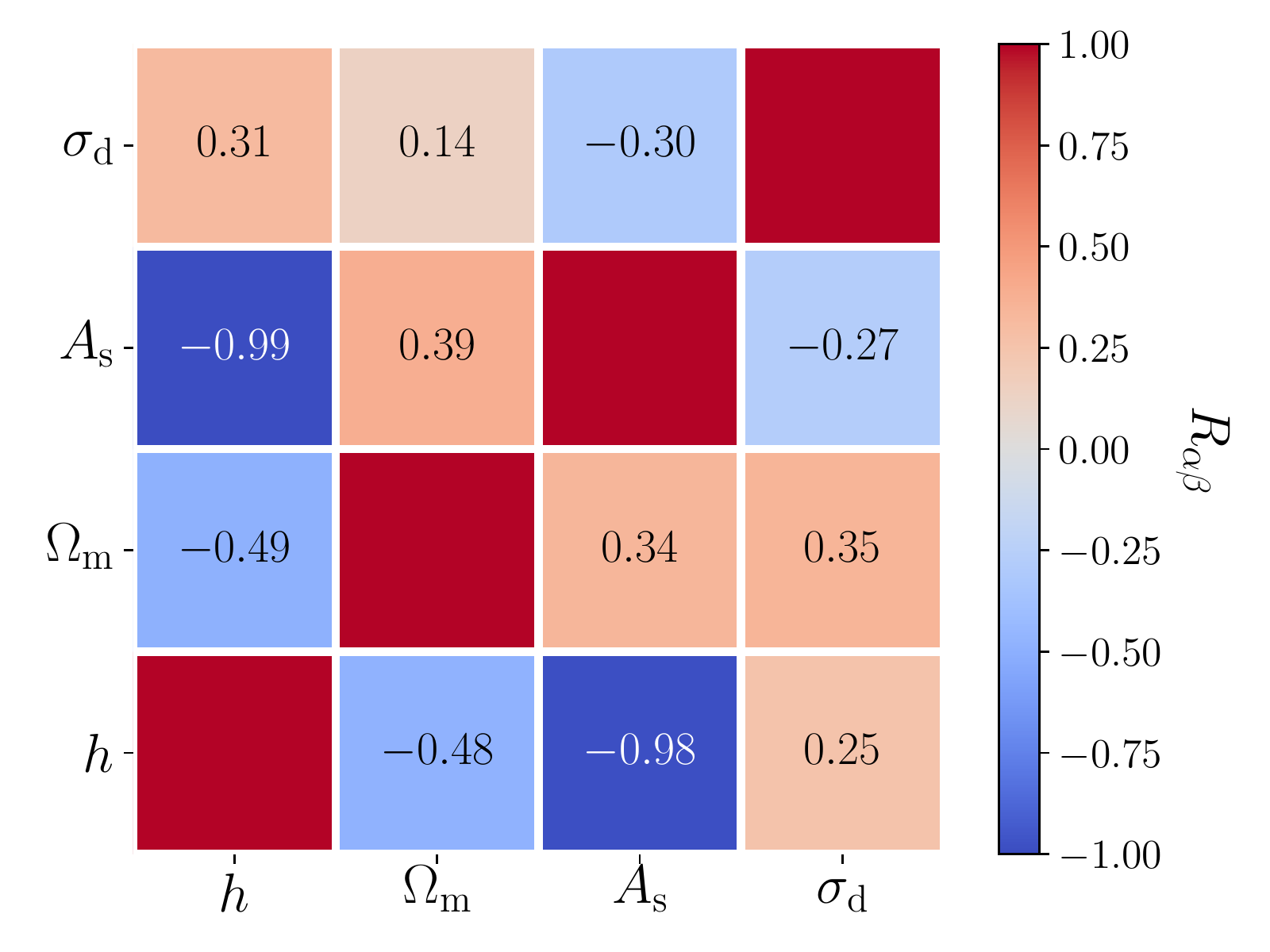}
  \caption{Correlation coefficients for \texttt{RegPT+} at 2-loop level.
  The upper left (lower right) triangle shows results
  with $k_\mathrm{max} = 0.18 \, (0.24) \, \hMpcinv$.
  The red (blue) parts correspond to positive (negative) correlations.}
  \label{fig:ccoef_sigma_d}
\end{figure}

\begin{figure}[htbp]
  \centering
  \includegraphics[width=0.45\textwidth]{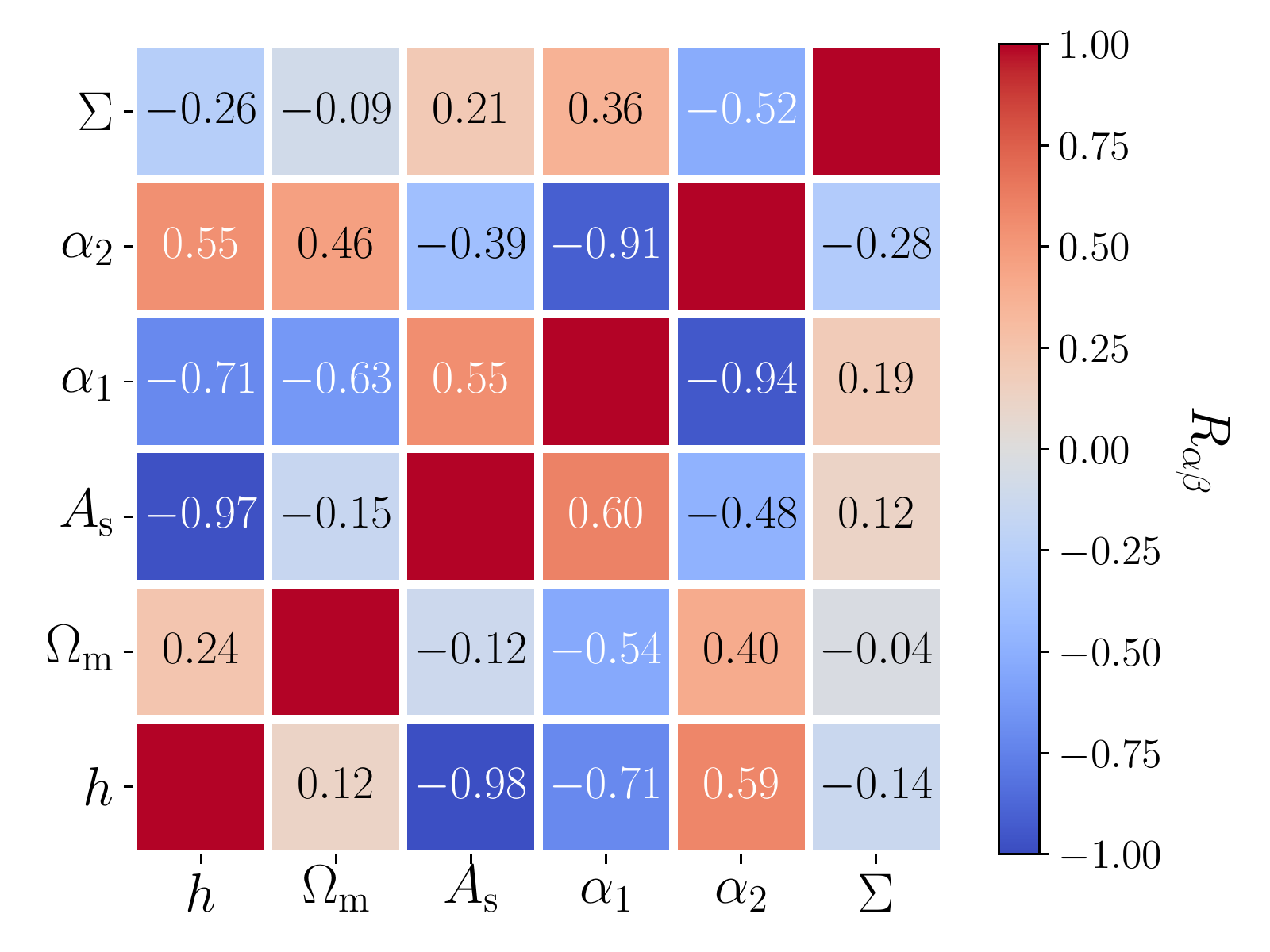}
  \caption{Correlation coefficients for IR-resummed EFT at 2-loop level.
  The upper left (lower right) triangle shows results
  with $k_\mathrm{max} = 0.18 \, (0.24) \, \hMpcinv$.
  The red (blue) parts correspond to positive (negative) correlations.}
  \label{fig:ccoef_IRresum}
\end{figure}

\begin{figure}[htbp]
  \centering
  \includegraphics[width=0.45\textwidth]{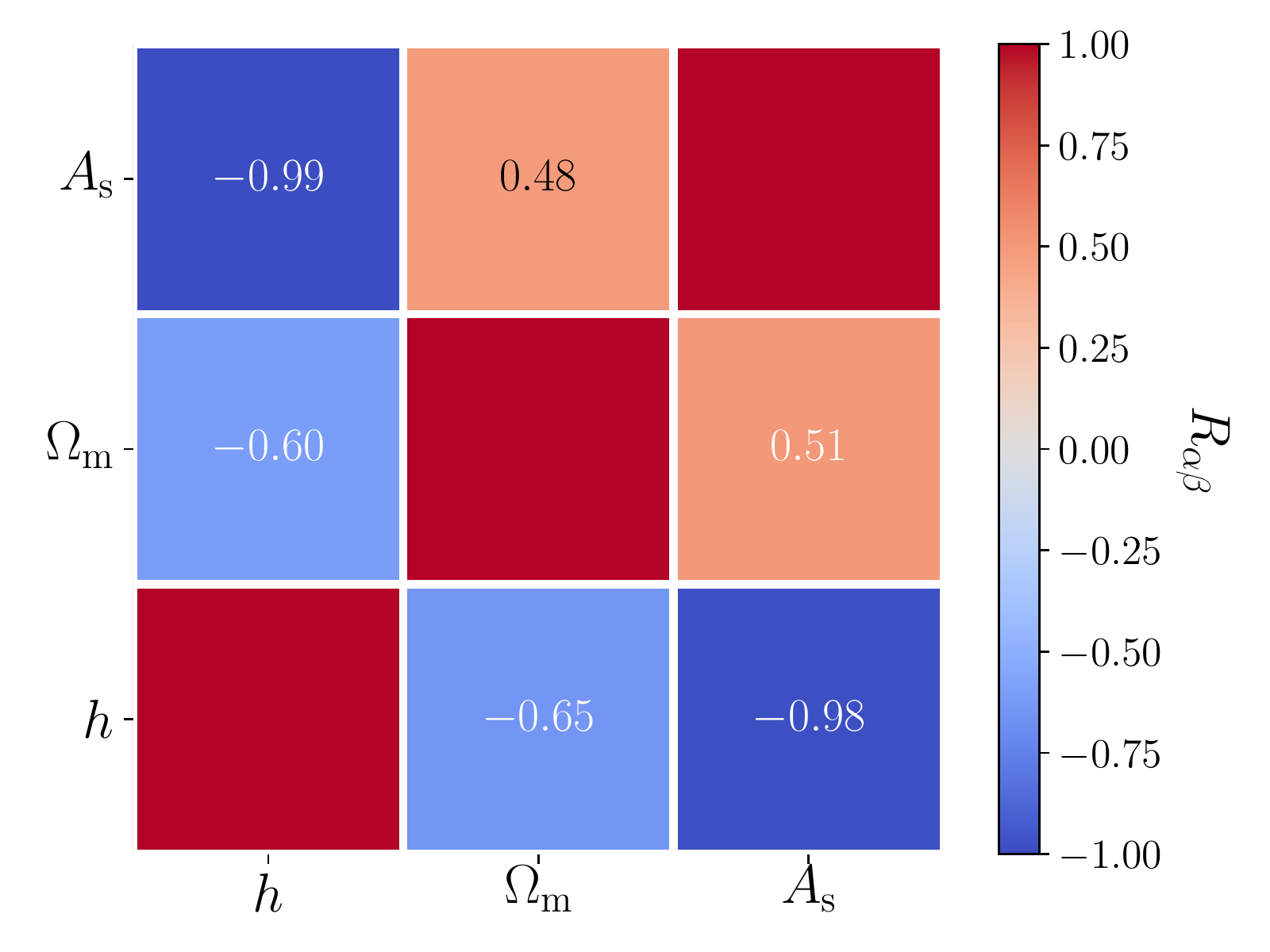}
  \caption{Correlation coefficients for \texttt{RESPRESSO}.
  The upper left (lower right) triangle shows results
  with $k_\mathrm{max} = 0.18 \, (0.24) \, \hMpcinv$.
  The red (blue) parts correspond to positive (negative) correlations.}
  \label{fig:ccoef_RESPRESSO}
\end{figure}

Generally, nuisance parameters help to improve fitting power spectrum even at small scales.
Simultaneously, if cosmological parameters are taken far from the true value,
nuisance parameters can adjust spectra and thus
constraints on cosmological parameters will be degraded.
This effect can be observed in the parameter degeneracy between cosmological and
nuisance parameters. The degeneracy means that the effect due to the cosmological parameter
can be compensated by changing the nuisance parameter.
On the other hand, when there is no degeneracy,
the nuisance parameter simply enhances the prediction capability of the model
or has almost no effects in the interested ranges.
In order to address this effect,
we quantify the degeneracy between parameters from
correlation coefficients defined as,
\beq
R_{\alpha \beta} \equiv
\frac{S_{\alpha \beta}}{\sqrt{S_{\alpha \alpha} S_{\beta \beta}}} .
\eeq
In Figs.~\ref{fig:ccoef_2loop}, \ref{fig:ccoef_SPT2loop},
\ref{fig:ccoef_sigma_d}, \ref{fig:ccoef_IRresum}, and \ref{fig:ccoef_RESPRESSO},
the correation coefficients for all pairs of parameters
with \texttt{RegPT}, SPT, \texttt{RegPT+}, IR-resummed EFT, and \texttt{RESPRESSO}
for $k_\mathrm{max} = 0.18, \, 0.24 \, \hMpcinv$ are shown.
We observe that changing $k_\mathrm{max}$ does not alter significantly
the structure of the correlation matrix.
However, if $k_\mathrm{max}$ is higher than the scale where each model is reliable
or equivalently the corresponding FoB is high, the correlations might be altered.
In \texttt{RegPT+} model, the nuisance parameter $\sigmad$ is
moderately degenerate with cosmological parameters,
and for IR-resummed EFT model, the parameters $\alpha_1$
and $\alpha_2$ are strongly correlated with cosmological parameters
though the parameter $\Sigma$ does not show strong correlation.
This degeneracy reduces the constraining power and leads to the suppression of FoM.
Since there exists stronger degeneracy in IR-resummed EFT model,
the suppression of FoM is more appreciable.
The parameter degeneracy also changes the degeneracy between
cosmological parameters, and the resultant correlation becomes
different from no nuisance parameter case via marginalization
(see Figs.~\ref{fig:ccoef_2loop} and \ref{fig:ccoef_SPT2loop}
in the cases of \texttt{RegPT} and SPT, respectively).

\section{Conclusions}
\label{sec:conclusions}
The measurement of matter power spectrum or
the BAO feature imprinted on it has been considered
to be one of the most fundamental observables in precision cosmology.
But in order to efficiently constrain cosmological models or parameters,
observables such as power spectrum should be
analytically modeled beyond the linear theory.
In this work, we explore the efficiency with which such models could constrain
the cosmological parameters focusing our analysis on
what could be derived from real space power spectrum.
The latter is obtained from an $N$-body simulation
for a specific set of cosmological parameters.
The performances of analytical models can then be scrutinized in terms of precision and accuracy.

The analytical approaches we employ are
\begin{itemize}
  \item Standard Perturbation Theory, (SPT),
  based on a direct expansion of the Euler fluid equations with respect to linear density contrast;
  \item Regularized Perturbation Theory, (\texttt{RegPT}),
  based on a reorganization of the series expansion with the help of the multipoint propagators;
  \item An extension of \texttt{RegPT} (\texttt{RegPT+})
  in which the damping scale is taken as a free parameter
  to account for the fact that it not predictable from first principle calculations;
  \item An IR-resummed Effective Field Theory (EFT) model
  in which non-PT parameters are introduced
  to account for the impact of the small scale physics on the growth of spectra,
  such as the effective pressure, etc.
\end{itemize}

Those models, all considered up to 2-loop order,
are more precisely described in Sec.~\ref{sec:theory}.
Our whole procedure is checked and calibrated
with the help of \texttt{RESPRESSO} which can accurately predict
how nonlinear power spectrum are deformed,
in the whole range of modes of interest,
when the cosmological parameters are varied.
Taking advantage of a simulated mock observation of real space power spectrum,
cosmological parameters are then fitted using the analytical models
described above with the MCMC technique.

In order to precisely quantify the performances of the codes we used the Figure of Bias (FoB),
which is the difference between estimated and input parameters
normalized by the variances, and the Figure of Merit (FoM),
which roughly corresponds to the inverse of the volume of the confidence region.
Our findings from the analysis with power spectrum at the redshift $z = 1$
can be summarized as follows,
\begin{itemize}
  \item \texttt{RegPT} and SPT give unbiased estimates of the cosmological parameters
  when the range of modes used to fit the parameters is
  limited to $k_\mathrm{max} = 0.24 \, \hMpcinv$. They fail for higher $k_\mathrm{max}$.
  \texttt{RegPT+} and IR-resummed EFT are accurate up to $k_\mathrm{max} = 0.30 \, \hMpcinv$
  thanks to the extra degrees of freedom they contain.
  \item On the other hand, as expected,
  FoMs of a given model monotonically increase as a function of $k_\mathrm{max}$
  as the amount of available information increases.
  And as expected the resulting precision in the cosmological parameters
  is all the more sensitive to $k_\mathrm{max}$
  that the number of useful modes scales like $k_\mathrm{max}^3$.
  \item FoMs of models with extra free parameters,
  \texttt{RegPT+} and IR-resummed EFT, are significantly reduced,
  by a factor respectively about 2 and 10,
  compared with those from models that are entirely predictive
  such as SPT and \texttt{RegPT} in our case.
  This effect is roughly independent of $k_\mathrm{max}$.
\end{itemize}

In order to address more precisely the origin of the latter reduction,
we investigate how nuisance parameters correlate with cosmological parameters
taking advantage of the fact that such correlation coefficients
can be coherently extracted from the MCMC chains.
Our results are presented in Sec.~\ref{sec:correlations}.
We find that some of the nuisance parameters are degenerate
with cosmological parameters degrading
the precision with which the latter are determined.
This effect is all the more important that the number of free parameters is large.
It is on the other hand quite independent of $k_\mathrm{max}$.
We are then put in a situation where a trade-off should be found between accuracy,
which calls for more free parameters, and precision, which calls for less.
The best performing prescription of those
we consider here is the \texttt{RegPT+} model,
having only one free nuisance parameter
while still being able to provide unbiased estimates
up to $k_\mathrm{max}$ about $0.33 \, \hMpcinv$.
This is unlikely however to be a definitive results:
other prescriptions could probably be as effective
and this validity range depends in effect
on the detailed covariance properties of the mock data.

Note that in this exercise,
we were forced to restrict the chains to a three dimensional cosmological parameter space
(six dimensional space in total for EFT models).
The reason is that the code we used here was not fast enough
to cope with larger dimensional space \footnote{For example, for \texttt{RegPT}
with $k_\mathrm{max} = 0.21 \, \hMpcinv$,
the length of converged MCMC chains is 132,000.
And each step roughly takes 1 minute with Intel Xeon E5-2695 v4 ($2.1 \, \mathrm{GHz}$).}.
Needless is to say that the larger the number of parameters is,
the slower the convergence of the MCMC procedure is.
For exploring larger parameter spaces,
it is then crucial to implement fast methods for computing power spectrum.
Several methods have already been proposed to speed up 2-loop level calculations
\cite{Schmittfull16a,Schmittfull16b,Simonovic18}.
Other aspects which should eventually be incorporated are modified gravity,
which is addressed in the context of EFT in Ref.~\cite{Bose18},
galaxy bias, and redshift space distortions effects.
They also lead to a further increase of the number of parameters to use.
These will be subjects of subsequent papers.

We are planning to release a set of numerical codes
to compute the power spectrum perturbatively
based on the fast scheme originally proposed by Ref.~\cite{Taruya12}.
The code suite will handle redshift space distortions and galaxy bias
with the fast scheme consistently up to the 2-loop level.
We have made an initial version of this code available on the repository
(\url{https://github.com/0satoken/Eclairs}).
The current version supports computation of matter power spectrum in real space
based on analytical approaches presented in this paper.
The codes are written in C++ with the python wrapper,
which is designed to be easily combined with MCMC samplers.

\begin{acknowledgments}
K.O. acknowledges helpful discussions with Masahiro Takada and Teppei Okumura.
K.O. is supported by Research Fellowships of the Japan Society for the
Promotion of Science (JSPS) for Young Scientists,
JSPS Overseas Challenge Program for Young Researchers
and Advanced Leading Graduate Course for Photon Science.
KO acknowledges the hospitality of Institut d'Astrophysique de Paris
where this work was initiated.
This work was supported by JSPS Grant-in-Aid for JSPS Research Fellow
Grant Number JP16J01512 (K.O.) and JSPS KAKENHI Grant Numbers JP17K14273 (T.N.),
JP15H05899 (A.T.), and JP16H03977 (A.T.).
T.N. also acknowledges financial support from
Japan Science and Technology Agency (JST) CREST Grant Number JPMJCR1414.
Numerical simulations were carried out on Cray XC50
at the Center for Computational Astrophysics,
National Astronomical Observatory of Japan
and Cray XC40 at Yukawa Institute Computer Facility, Kyoto University.
\end{acknowledgments}

\appendix
\section{Explicit formulas of analytical approaches}
In this Appendix, we present explicit formulas for power spectrum
based on SPT, \texttt{RegPT}, and IR-resummed EFT.

\subsection{SPT}
\label{sec:ex_SPT}
The correction terms of power spectrum based on SPT
at 1-loop and 2-loop levels are
\beqa
\Delta P^{\text{SPT}}_{\text{1-loop}} (k) &=& D_+^4 [ 2 P_{13} (k) + P_{22} (k) ] ,
\label{eq:SPT1loop} \\
\Delta P^{\text{SPT}}_{\text{2-loop}} (k) &=& D_+^6 [ 2 P_{15} (k) + 2 P_{24} (k) + P_{33} (k) ].
\label{eq:SPT2loop}
\eeqa
Each correction term is given as
\begin{widetext}
\beqa
P_{13} (k) &=& 3 P_0 (k) \int \frac{d^3 q}{(2\pi)^3}
F_\mathrm{sym}^{(3)} (\bm{k}, \bm{q}, -\bm{q}) P_0 (q) , \\
P_{22} (k) &=& 2 \int \frac{d^3 q}{(2\pi)^3}
[F_\mathrm{sym}^{(2)} (\bm{q}, \bm{k} - \bm{q}) ]^2
P_0 (q) P_0 (|\bm{k} - \bm{q}|), \\
P_{15} (k) &=& 15 P_0 (k)
\int \frac{d^3 q_1}{(2\pi)^3} \frac{d^3 q_2}{(2\pi)^3}
F_\mathrm{sym}^{(5)} (\bm{k}, \bm{q}_1, -\bm{q}_1, \bm{q}_2, -\bm{q}_2)
P_0 (q_1) P_0 (q_2) , \\
P_{24} (k) &=& 12 \int \frac{d^3 q_1}{(2\pi)^3} \frac{d^3 q_2}{(2\pi)^3}
F_\mathrm{sym}^{(2)} (\bm{q}_1, \bm{k} - \bm{q}_1)
F_\mathrm{sym}^{(4)} (\bm{q}_1, \bm{k} - \bm{q}_1, \bm{q}_2, -\bm{q}_2)
P_0 (q_1) P_0 (q_2) P_0 (|\bm{k} - \bm{q}_1|) , \\
P_{33} (k) &=& 9 P_0 (k) \left[ \int \frac{d^3 q}{(2\pi)^3}
F_\mathrm{sym}^{(3)} (\bm{k}, \bm{q}, -\bm{q}) P_0 (q) \right]^2 \nonumber \\
& & + 6 \int \frac{d^3 q_1}{(2\pi)^3} \frac{d^3 q_2}{(2\pi)^3}
[F_\mathrm{sym}^{(3)} (\bm{q}_1, \bm{q}_2, \bm{k} - \bm{q}_1 - \bm{q}_2)]^2
P_0 (q_1) P_0 (q_2) P_0 (|\bm{k} - \bm{q}_1 - \bm{q}_2|) .
\eeqa
\end{widetext}
Eventually, the power spectra at 1-loop and 2-loop levels are given as
\beqa
P^{\text{SPT}}_{\oneloop} (k) &=& P_\Lrm (k) +
\Delta P^{\text{SPT}}_{\text{1-loop}} (k) ,\\
P^{\text{SPT}}_{\twoloop} (k) &=& P_\Lrm (k) +
\Delta P^{\text{SPT}}_{\text{1-loop}} (k) +
\Delta P^{\text{SPT}}_{\text{2-loop}} (k) .
\eeqa

\subsection{\texttt{RegPT}}
\label{sec:ex_RegPT}
The expression of power spectrum of \texttt{RegPT} at 2-loop level should be
\begin{widetext}
\beqa
P^\texttt{RegPT}_{\text{2-loop}} (k) &=& [ \Gamma_\mathrm{reg}^{(1)} (k) ]^2 P_0 (k) +
2 \int \frac{d^3 q}{(2 \pi)^3} [\Gamma_\mathrm{reg}^{(2)}
(\bm{q}, \bm{k}-\bm{q})]^2 P_0 (q) P_0 (|\bm{k}-\bm{q}|)
\nonumber \\
&& + 6 \int \frac{d^3 q_1}{(2 \pi)^3} \frac{d^3 q_2}{(2 \pi)^3} [\Gamma_\mathrm{reg}^{(3)}
(\bm{q}_1, \bm{q}_2, \bm{k}-\bm{q}_1-\bm{q}_2)]^2
P_0 (q_1) P_0 (q_2) P_0 (|\bm{k}-\bm{q}_1-\bm{q}_2|) ,
\eeqa
\end{widetext}
where the regularized propagators are expressed as,
\beqa
\Gamma^{(1)}_\mathrm{reg} (k) &=& D_+ \left[ 1 + \alpha_k + \frac{1}{2} \alpha_k^2
\right. \nonumber \\
&& \left. + D_+^2 \Gamma^{(1)}_\oneloop (k) (1+\alpha_k)
+ D_+^4 \Gamma^{(1)}_\twoloop (k) \right] \nonumber \\
&& \times \exp(-\alpha_k) , \\
\Gamma^{(2)}_\mathrm{reg} (\bm{k}_1, \bm{k}_2) &=& D_+^2 \left[ (1 + \alpha_k)
F^{(2)}_\mathrm{sym} (\bm{k}_1, \bm{k}_2) \right. \nonumber \\
&& \left. + D_+^2 \Gamma^{(2)}_\oneloop (\bm{k}_1, \bm{k}_2) \right] \exp(-\alpha_k) , \\
\Gamma^{(3)}_\mathrm{reg} (\bm{k}_1, \bm{k}_2, \bm{k}_3) &=&
D_+^3 F^{(3)}_\mathrm{sym} (\bm{k}_1, \bm{k}_2, \bm{k}_3) \exp(-\alpha_k) ,
\eeqa
where $\alpha_k = (1/2) k^2 D_+^2 \sigmad^2$, and $\Gamma^{(1)}_\oneloop (k)$,
$\Gamma^{(1)}_\twoloop (k)$, and $\Gamma^{(2)}_\oneloop (\bm{k}_1, \bm{k}_2)$ are
defined in Eq.~\eqref{eq:gamma_propagator}.

The power spectrum at 1-loop level is given as
\beqa
P^\texttt{RegPT}_{\text{1-loop}} (k) &=& [ \Gamma_\mathrm{reg}^{(1)} (k) ]^2 P_0 (k)
\nonumber \\
&& + 2 \int \frac{d^3 q}{(2 \pi)^3} [\Gamma_\mathrm{reg}^{(2)}
(\bm{q}, \bm{k}-\bm{q})]^2 P_0 (q) P_0 (|\bm{k}-\bm{q}|) ,
\nonumber \\
\eeqa
with the corresponding regularized propagators $\Gamma^{(1)}_\mathrm{reg}$
and $\Gamma^{(2)}_\mathrm{reg}$,
\beqa
\Gamma^{(1)}_\mathrm{reg} (k) &=& D_+ \left[ 1 + \alpha_k +
D_+^2 \Gamma^{(1)}_\oneloop (k) \right] \nonumber \\
& & \times \exp(-\alpha_k) , \\
\Gamma^{(2)}_\mathrm{reg} (\bm{k}_1, \bm{k}_2) &=&
D_+^2 F^{(2)}_\mathrm{sym} (\bm{k}_1, \bm{k}_2) \exp(-\alpha_k) .
\eeqa

\subsection{IR-resummed EFT}
\label{sec:ex_IRresum}
In the following, we give expressions for matter power spectrum
at 2-loop and 1-loop levels based on the IR-resummed EFT approach.
At 2-loop level, the matter power spectrum is given as,
\begin{widetext}
\beqa
P^\text{IR EFT}_\twoloop (k) &=& P^\nwrm (k) + P^\wrm (k) ; \\
P^\nwrm (k) &=& (1+\alpha_1 k^2) P_\Lrm^\nwrm (k) + (1+\alpha_2 k^2) \Delta P_\oneloop^\nwrm (k)
+\Delta P^\nwrm_\twoloop (k) , \\
P^\wrm (k) &=& e^{-k^2 \Sigma^2} \left[ (1+\alpha_1 k^2 + C_1) P_\Lrm^\wrm (k)
+ (1+\alpha_2 k^2 + C_2) \Delta P_\oneloop^\wrm (k) + \Delta P_\twoloop^\wrm (k) \right],
\eeqa
\end{widetext}
where
\beqa
C_1 &=& k^2 \Sigma^2 (1+\alpha_1 k^2) + \frac{1}{2} k^4 \Sigma^4 , \\
C_2 &=& k^2 \Sigma^2 (1+\alpha_2 k^2) .
\eeqa
Here, we have introduced three free parameters $\alpha_1$, $\alpha_2$, and
$\Sigma$, which are usually calibrated with $N$-body simulations.
For no-wiggle linear spectra $P_\Lrm^\nwrm (k)$,
we smooth linear power spectrum as described in Eqs.~\eqref{eq:smoothing1}
and \eqref{eq:smoothing2}, i.e.,
\beqa
P_\Lrm^\nwrm (k) &=& P_\mathrm{EH} (k)
\frac{1}{\sqrt{2\pi} \log_{10} \lambda} \int d(\log_{10} q)
\frac{P_\Lrm (q)}{P_\mathrm{EH} (q)}
\nonumber \\
&& \times \exp \left[ -\frac{(\log_{10} k - \log_{10} q)^2}{2 (\log_{10} \lambda)^2} \right] ,
\eeqa
where we adopt the smoothing scale as $\lambda = 10^{0.25} \, \hMpcinv$ and
$P_\mathrm{EH} (k)$ is the power spectrum without wiggle feature \cite{Eisenstein98}.
The residual corresponds to the wiggle part $P_\Lrm^\wrm (k)$, i.e.,
\beq
P_\Lrm^\wrm (k) = P_\Lrm (k) - P_\Lrm^\nwrm (k) .
\eeq
Then, we plug the smoothed spectrum $P_\Lrm^\nwrm (k)$, instead of linear spectrum $P_\Lrm (k)$,
into the SPT formulas (Eqs.~\ref{eq:SPT1loop} and \ref{eq:SPT2loop})
to obtain no-wiggle correction terms at 1-loop and 2-loop levels
($\Delta P_\oneloop^\nwrm$ and $\Delta P_\twoloop^\nwrm$).
The wiggle parts of correction terms are also obtained as residuals,
\beqa
\Delta P_\oneloop^\wrm (k) &=& \Delta P^{\text{SPT}}_{\oneloop} (k) -
\Delta P_\oneloop^\nwrm (k) , \\
\Delta P_\twoloop^\wrm (k) &=& \Delta P^{\text{SPT}}_{\twoloop} (k) -
\Delta P_\twoloop^\nwrm (k) .
\eeqa

For 1-loop level, the matter power spectrum is given as,
\beqa
P^\text{IR EFT}_\oneloop(k) &=& P^\nwrm (k) + P^\wrm (k) , \\
P^\nwrm (k) &=& (1+\alpha_1 k^2) P_\Lrm^\nwrm (k) + \Delta P_\oneloop^\nwrm (k) ,
\nonumber \\
&& \\
P^\wrm (k) &=& e^{-k^2 \Sigma^2} \left[ (1+\alpha_1 k^2 + k^2 \Sigma^2)
P_\Lrm^\wrm (k) \right. \nonumber \\
&& \left. + \Delta P_\oneloop^\wrm (k) \right] .
\eeqa
In this case, there are two free parameters, $\alpha_1$ and $\Sigma$,
which are also calibrated against $N$-body simulations.

\subsection{EFT}
\label{sec:ex_EFT}
We have considered IR-resummed EFT so far but there is
a simpler description of EFT.
For 1-loop level, we can write down the expression for EFT as,
\beq
P^\text{EFT}_{\oneloop} (k) = P^\text{SPT}_{\oneloop} (k)
- 2 (2 \pi) c_{s(1)}^2 \left( \frac{k}{k_\mathrm{NL}} \right)^2 P_\Lrm (k) ,
\eeq
where $k_\mathrm{NL}$ is the nonlinear scale and
$c_{s(1)}$ is the effective sound speed \cite{Carrasco14}.
We treat the combination $c_{s(1)}/ k_\mathrm{NL}$ as a free parameter.
The EFT prescription is based on the similar idea for IR-resummed EFT,
and roughly speaking the parameter $\alpha_1$ in IR-resummed EFT corresponds to
the sound speed in EFT.
Thus, these two models should provide us with similar results.
However, IR-resummed EFT introduces another free parameter $\Sigma$ which regulates
the damping feature at small scales
and when $\Sigma = 0$, the expression can be reduced to that of EFT.

\section{Results with power spectra at 1-loop level}
\label{sec:1loop_results}
\begin{figure*}[htbp]
  \centering
  \includegraphics[width=0.9\textwidth]{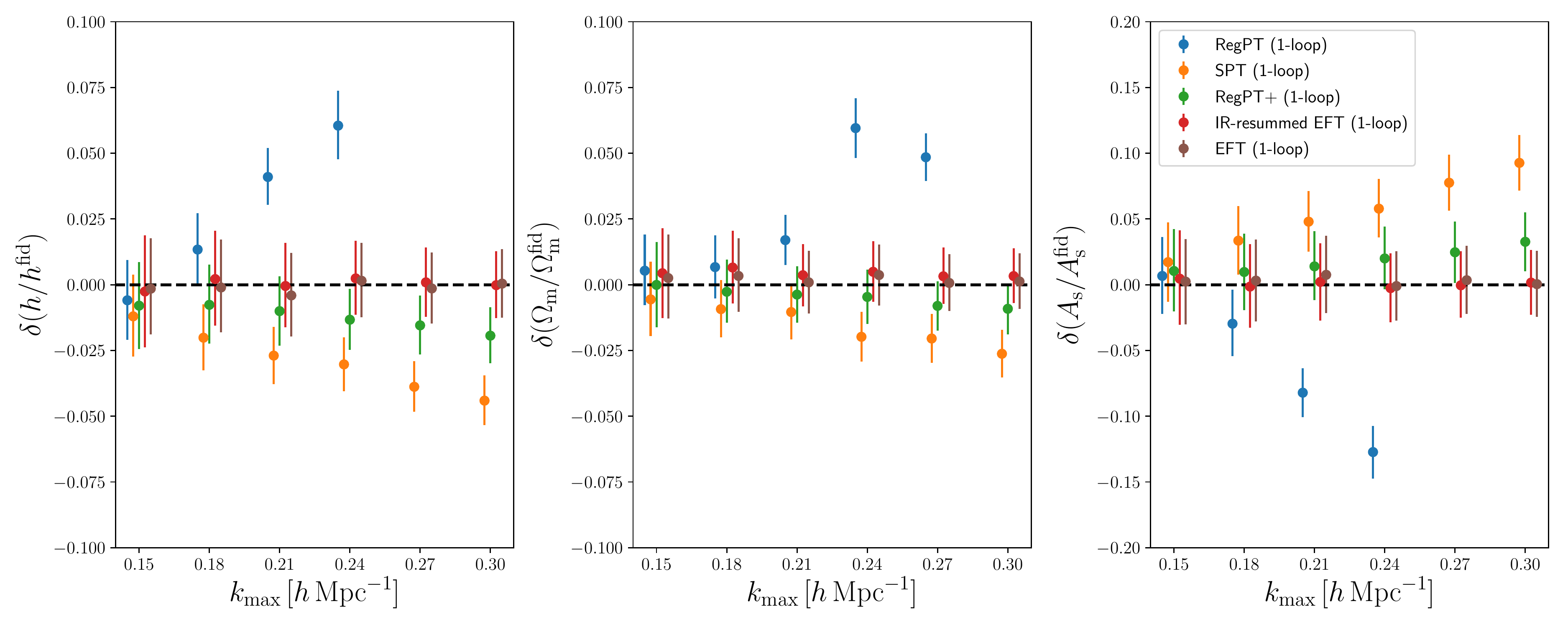}
  \caption{Medians of cosmological parameters estimated from MCMC chains
  with errors as the fractional ratios
  with respect to the fiducial values.
  All of estimates are based on 1-loop level calculations.
  The lower (upper) limit of error bars
  correspond to $16 \%$ ($84 \%$) percentile.}
  \label{fig:param_1loop}
\end{figure*}

\begin{figure*}[htbp]
  \centering
  \includegraphics[width=0.9\textwidth]{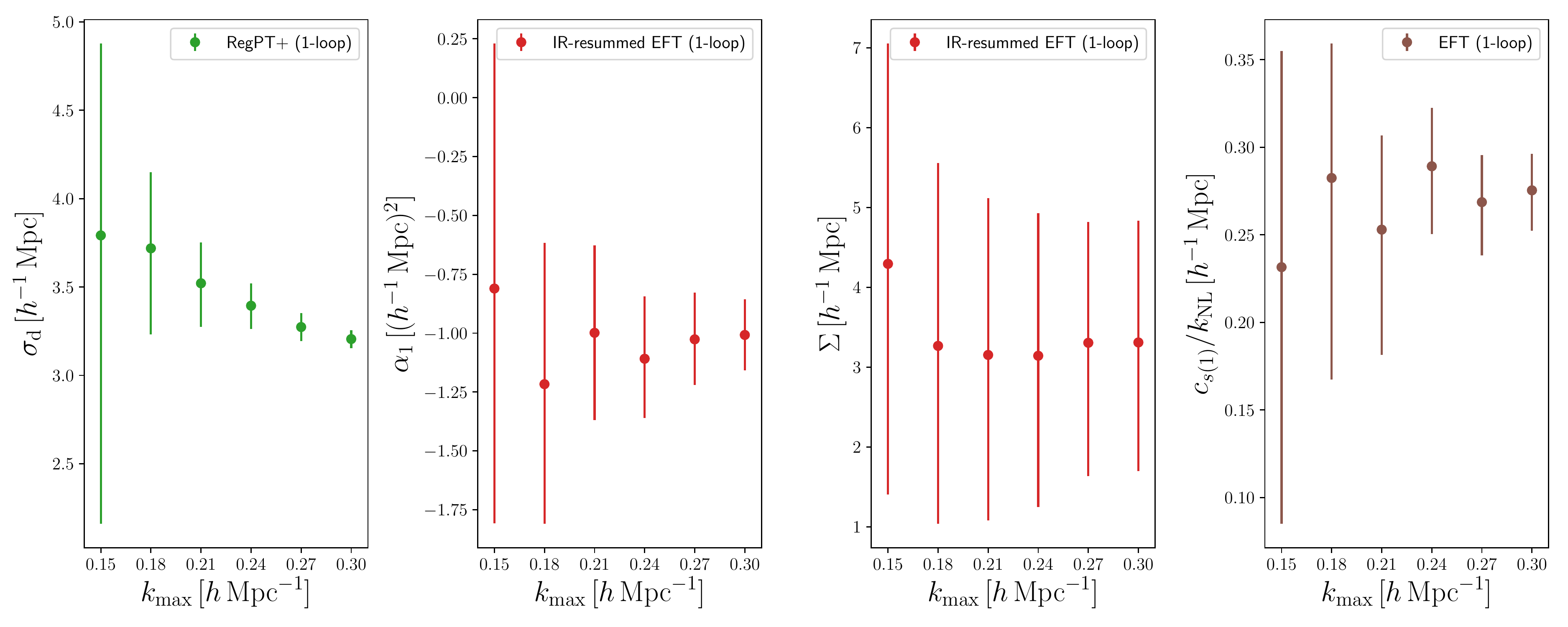}
  \caption{Medians of nuisance parameters, $\sigmad$ for \texttt{RegPT+},
  $\alpha_1$ and $\Sigma$ for IR-resummed EFT, and $c_{s(1)} / k_\mathrm{NL}$ for EFT,
  estimated from MCMC chains.
  The lower (upper) limit of error bars
  correspond to $16 \%$ ($84 \%$) percentile.}
  \label{fig:nuisance_1loop}
\end{figure*}

\begin{figure}[htbp]
  \centering
  \includegraphics[width=0.45\textwidth]{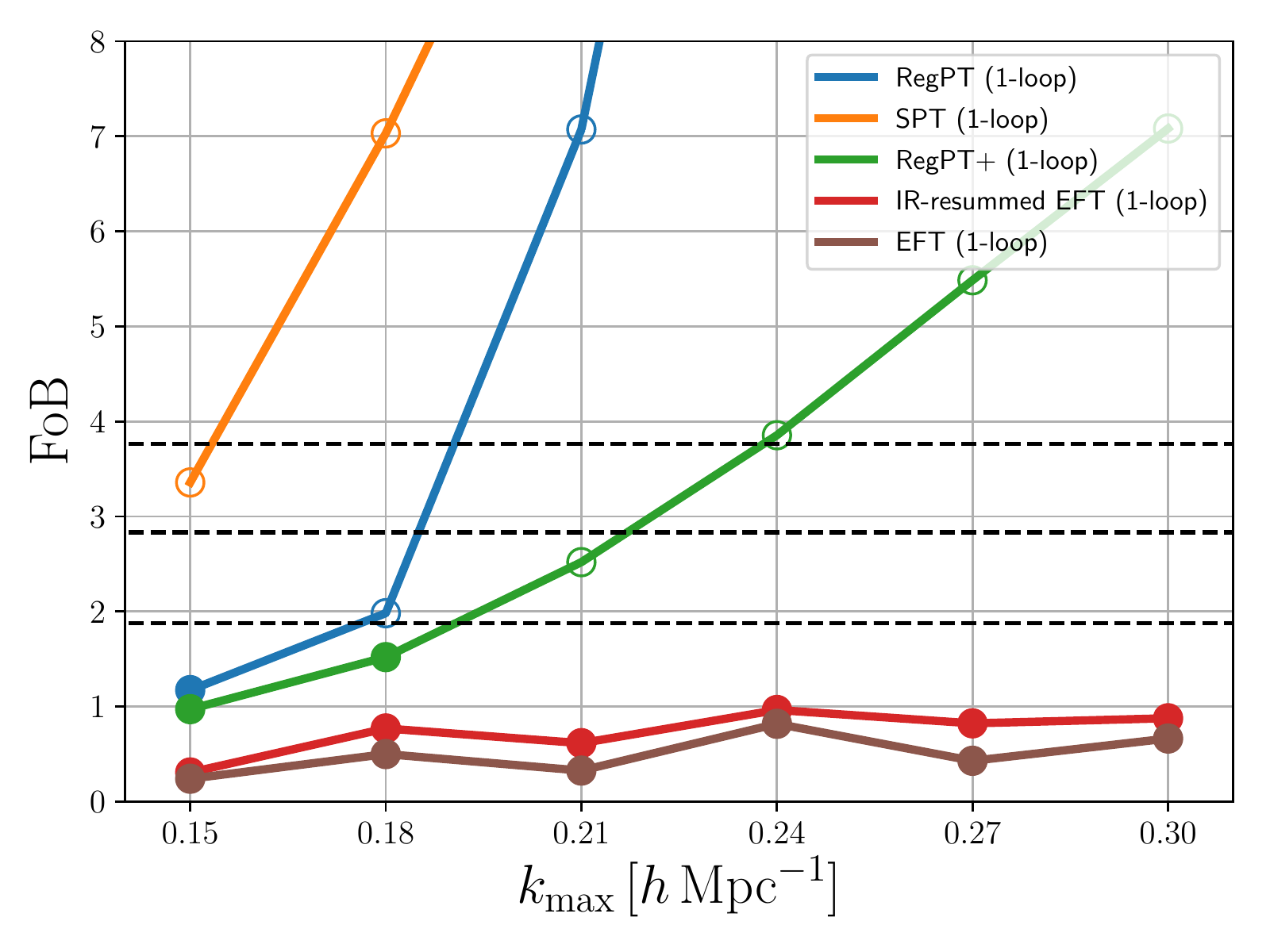}
  \caption{Figure of bias for different models at at 1-loop level
  estimated from MCMC chains.
  The black dashed line shows the $1\text{-}\sigma$, $2\text{-}\sigma$, and
  $3\text{-}\sigma$ critical values $1.88$, $2.83$, and $3.76$.
  The open (filled) symbols represent that the figure of bias
  exceeds (falls below) the $1\text{-}\sigma$ critical value.}
  \label{fig:bias_1loop}
\end{figure}

\begin{figure}[htbp]
  \centering
  \includegraphics[width=0.45\textwidth]{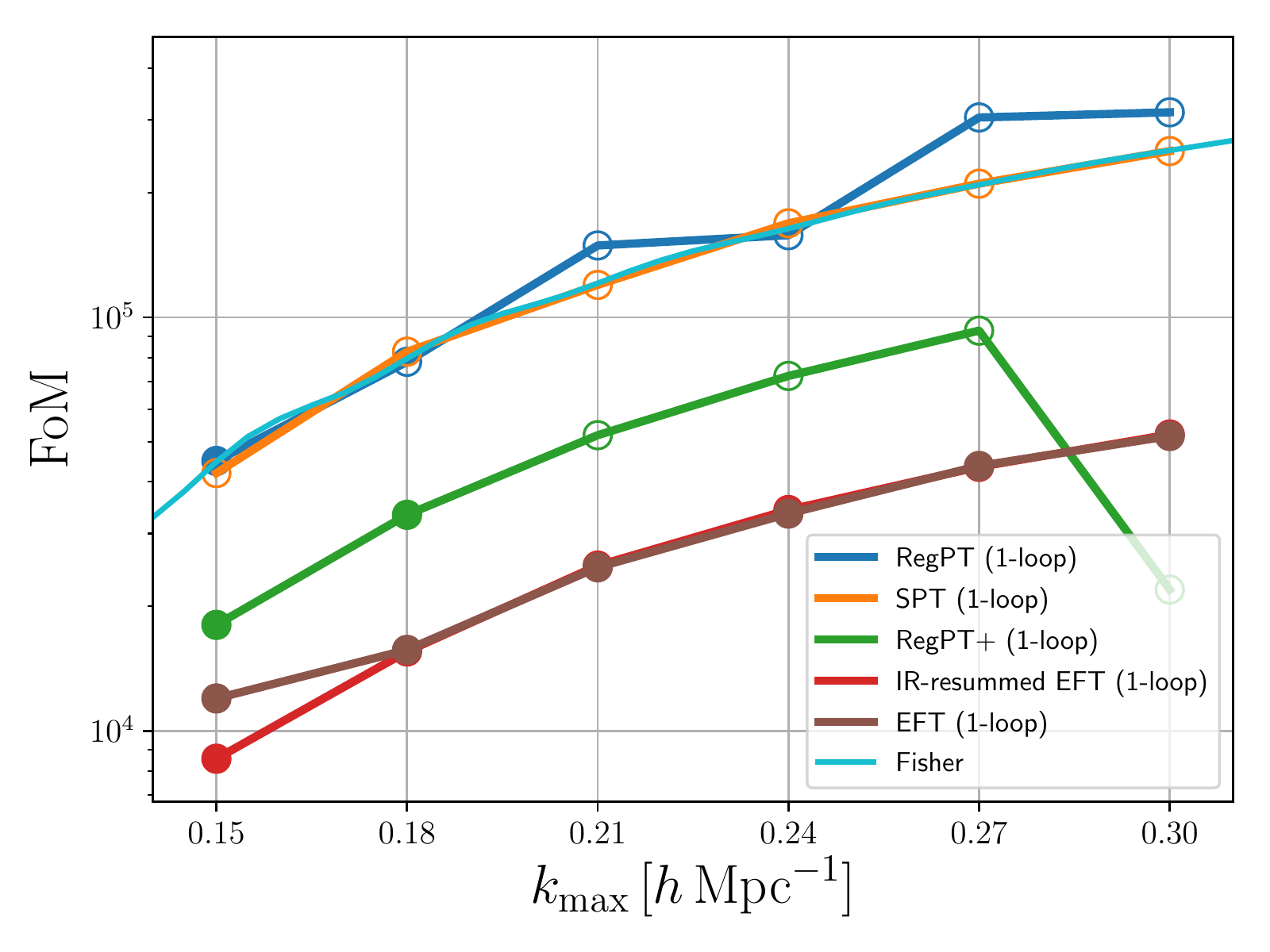}
  \caption{Figure of merit for different models at 1-loop level
  estimated from MCMC chains.
  The cyan line shows figure of merit from Fisher matrix with \texttt{RESPRESSO}.
  The open (filled) symbols represent that the corresponding figure of bias
  exceeds (falls below) the $1\text{-}\sigma$ critical value (see Fig.~\ref{fig:bias_1loop}).}
  \label{fig:merit_1loop}
\end{figure}

\begin{figure}[htbp]
  \centering
  \includegraphics[width=0.45\textwidth]{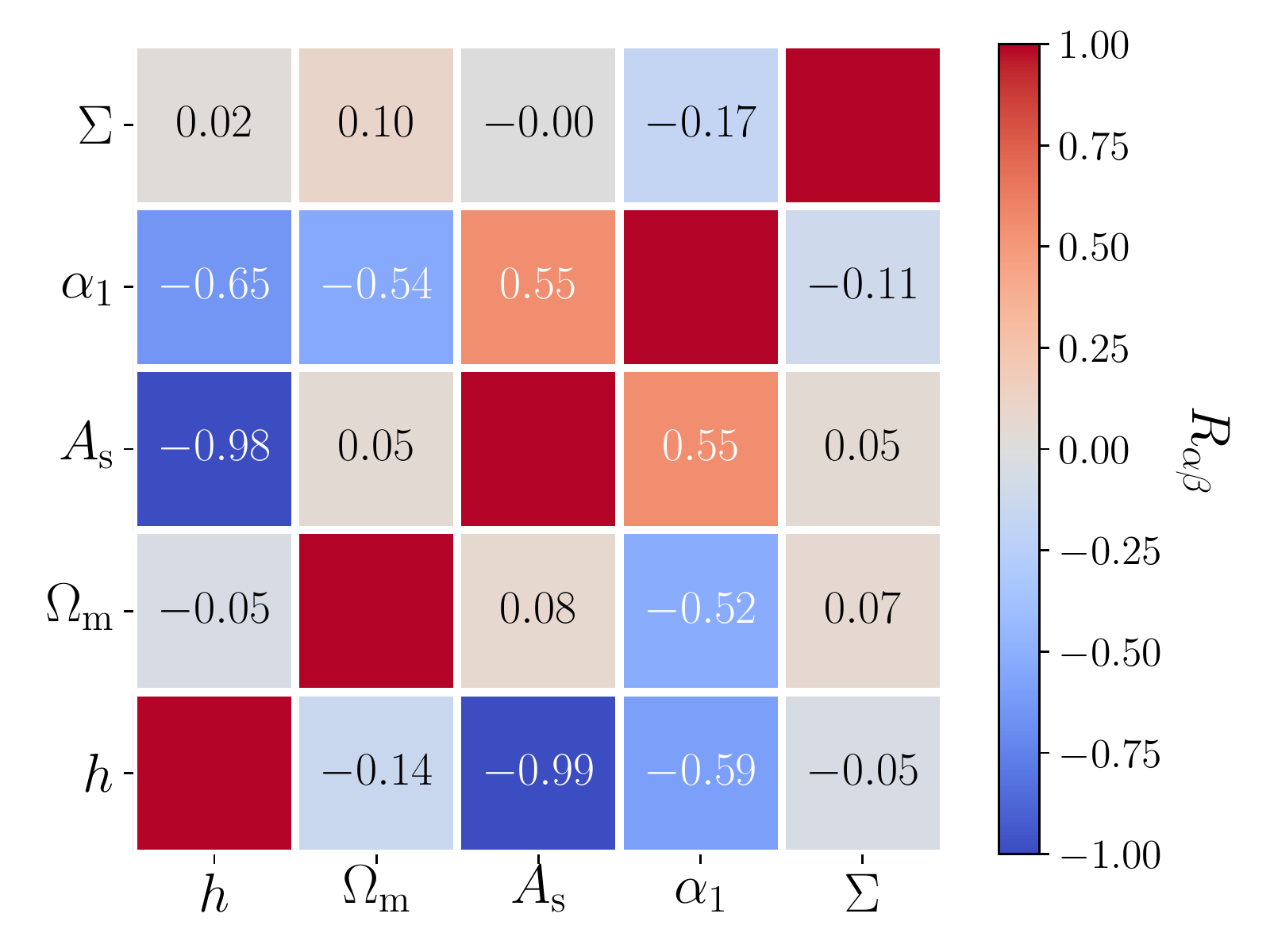}
  \caption{Correlation coefficients for IR-resummed EFT model at 1-loop level.
  The upper left (lower right) triangle shows results
  with $k_\mathrm{max} = 0.18 \, (0.24) \, \hMpcinv$.
  The red (blue) parts correspond to positive (negative) correlations.}
  \label{fig:ccoef_IRresum_1loop}
\end{figure}

\begin{figure}[htbp]
  \centering
  \includegraphics[width=0.45\textwidth]{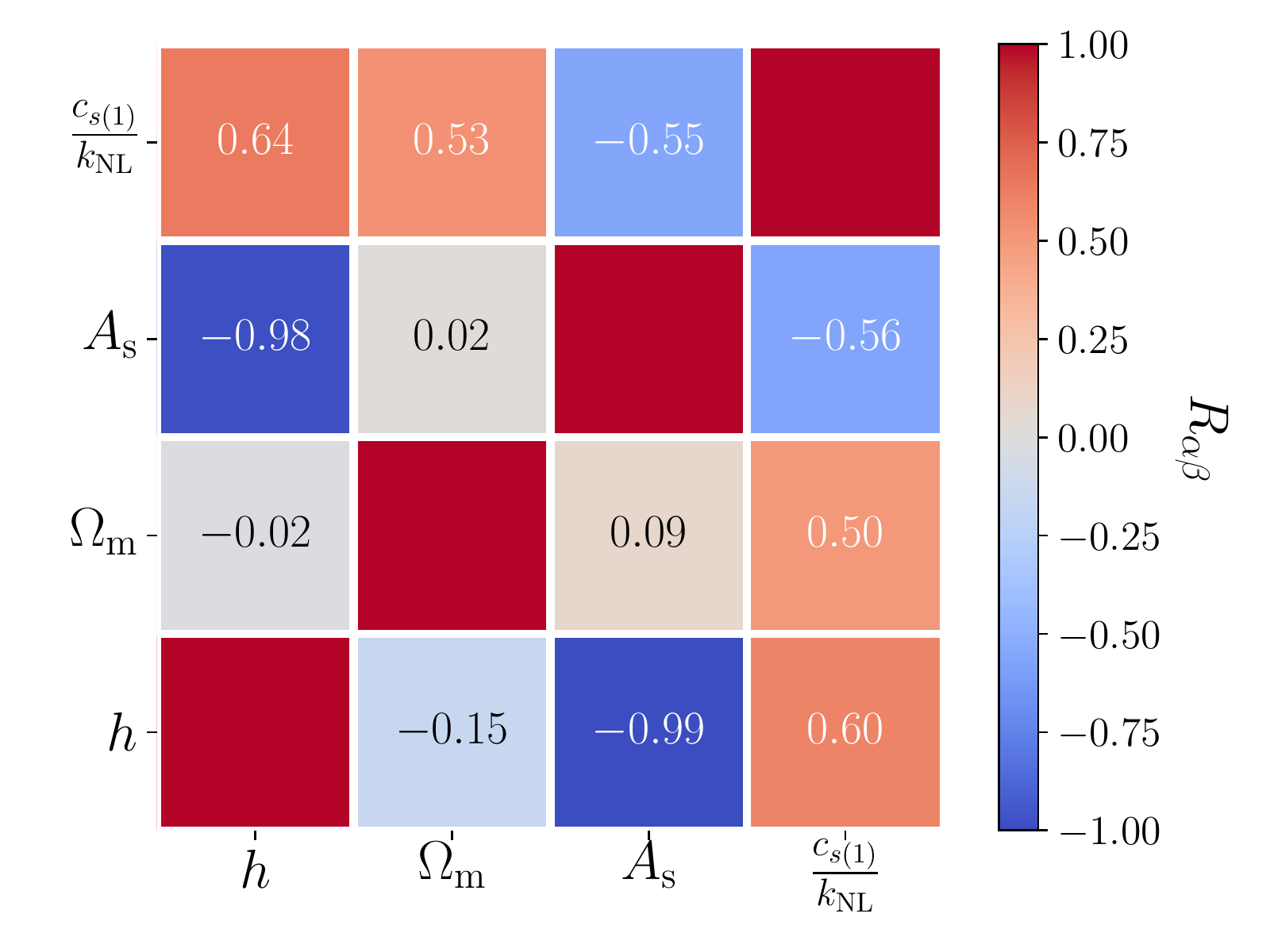}
  \caption{Correlation coefficients for EFT at 1-loop level.
  The upper left (lower right) triangle shows results
  with $k_\mathrm{max} = 0.18 \, (0.24) \, \hMpcinv$.
  The red (blue) parts correspond to positive (negative) correlations.}
  \label{fig:ccoef_EFT1loop}
\end{figure}

For comparison, we present results with methods at 1-loop level.
In Figs.~\ref{fig:param_1loop} and \ref{fig:nuisance_1loop},
cosmological and nuisance parameters estimated
with 1-loop level calculations are shown.
The results look mostly similar to 2-loop cases.
However, the estimated parameters start to deviate from true values
from smaller $k_\mathrm{max}$.
We can see a similar trend in FoB and FoM shown in
Figs.~\ref{fig:bias_1loop} and \ref{fig:merit_1loop}.
According to FoM, the constraining power is almost the same as
that at 2-loop level. On the other hand, in all cases FoBs are higher than
the counterpart in the 2-loop case.
The 2-loop level calculations outperform 1-loop calculations
though they are computationally more expensive.
The FoM of EFT is slightly larger than that of IR-resummed EFT but
the effect is subdominant.
In order to investigate the reason, we show correlation coefficients
for IR-resummed EFT and EFT in Figs.~\ref{fig:ccoef_IRresum_1loop}
and \ref{fig:ccoef_EFT1loop}.
In IR-resummed EFT model, the parameter $\Sigma$ has almost no
correlations with cosmological parameters.
That results in no degradation of parameter constraints due to $\Sigma$.
On the other hand, $\alpha_1$ in IR-resummed EFT and $c_{s(1)}/k_\mathrm{NL}$
in EFT correlates with cosmological parameters to similar extent, and thus
FoMs for these two models are almost the same.

\bibliography{main}

\end{document}